\documentclass[aps,prb,superscriptaddress,twocolumn,showpacs,floatfix,a4paper]{revtex4-1}

\usepackage{amsfonts,amsmath,amssymb}
\usepackage{textcomp}
\usepackage{bm}
\usepackage{upgreek}
\usepackage{graphicx}
\usepackage[latin1]{}
\usepackage{color}

\newcommand{\notop}{{{}_{}}}

\newcommand{\ie}{\textit{i.e.}}
\newcommand{\eg}{\textit{e.g.}}

\newcommand{\cf}{\textit{cf.}}

\renewcommand{\vec}[1]{\bm{#1}}
\newcommand{\ee}{\mathrm{e}}
\newcommand{\ii}{\mathrm{i}}

\newcommand{\fracsmall}[2]{\mbox{$\frac{#1}{#2}$}}

\newcommand{\aaa}{\vec{a}}

\newcommand{\GGG}{\vec{G}^\notop}

\newcommand{\kkk}{\vec{k}}

\newcommand{\rrr}{\vec{r}}

\newcommand{\VVV}{\vec{V}}

\newcommand{\calA}{\mathcal{A}}
\newcommand{\calQ}{\mathcal{Q}}

\newcommand{\calT}{\mathcal{T}}

\newcommand{\eps}{\epsilon}

\newcommand{\beq}[1]{\begin{equation} \eqlab{#1}}
\newcommand{\eeq}{\end{equation}}
\newcommand{\bsub}{\begin{subequations}}
\newcommand{\esub}{\end{subequations}}
\def\bal#1\eal{\begin{align}#1\end{align}}
\def\bsubal#1\esubal{\bsub \begin{align}#1\end{align} \esub}

\newcommand{\eqlab}[1]{\label{eq:#1}}
\renewcommand{\eqref}[1]{Eq.~(\ref{eq:#1})}
\newcommand{\eqsref}[2]{Eqs.~(\ref{eq:#1}) and~(\ref{eq:#2})}

\newcommand{\figref}[1]{Fig.~\ref{fig:#1}}

\newcommand{\appref}[1]{Appendix~\ref{sec:#1}}

\newcommand{\secref}[1]{Section~\ref{sec:#1}}

\newcommand{\seclab}[1]{\label{sec:#1}}

\begin{document}

\title{Dual-probe spectroscopic fingerprints of defects in graphene}

\author{Mikkel Settnes}\email[]{mikse@nanotech.dtu.dk}
\author{Stephen R. Power}
\author{Dirch H. Petersen}
\author{Antti-Pekka Jauho}
\affiliation{Center for Nanostructured Graphene (CNG), Department of Micro and Nanotechnology, DTU Nanotech, Technical University of Denmark, DK-2800 Kongens Lyngby, Denmark}

\date{\today}

\begin{abstract}
Recent advances in experimental techniques emphasize the usefulness of multiple scanning probe techniques when analyzing nanoscale samples.
Here, we analyze theoretically dual-probe setups with probe separations in the nanometer range, \ie , in a regime where quantum coherence effects can be observed at low temperatures. In a dual-probe setup the electrons are injected at one probe and collected at the other. The measured conductance reflects the local {\em transport properties} on the nanoscale, thereby yielding information complementary to that obtained with a standard one-probe setup (the local density-of-states).
In this work we develop a real space Green's function method to compute the conductance. This requires an extension of the standard calculation schemes, which typically address a finite sample between the probes. In contrast, the developed method makes no assumption on the sample size (\eg , an extended graphene sheet). 
Applying this method, we study the transport anisotropies in pristine graphene sheets, and analyze the spectroscopic fingerprints arising from quantum interference around single-site defects, such as vacancies and adatoms.
Furthermore, we demonstrate that the dual-probe setup is a useful tool for characterizing the electronic transport properties of extended defects or designed nanostructures.
In particular, we show that nanoscale perforations, or antidots, in a graphene sheet display Fano-type resonances with a strong dependence on the edge geometry of the perforation.
\end{abstract}

\maketitle

\section{Introduction}
A key step towards developing novel applications for graphene and other two-dimensional materials \cite{Novoselov2005,Dean2010,Wang2012} is to obtain a detailed understanding of their electron transport properties on the nanoscale. \cite{CastroNeto2009}
At these length scales structural details play a crucial role due to the restricted dimensionality.
Thus, studying spatially resolved electron transport becomes important, especially near defects and boundaries, which dramatically affect the conductance of a device. \cite{Peres2006,Chen2008}

Scanning Tunnelling Microscopy (STM)\cite{Binnig1982,Deshpande2012} is an important non-invasive method for studying the electronic structure of surfaces.
Nanometer scale STM measurements, yielding both local density of states (LDOS) and topographic details, are extensively used both theoretically \cite{Cheianov2006,Bena2008,Pellegrino2009,Peres2009,Mark2012,Bergvall2013,Lawlor2013,Lounis2014} and experimentally \cite{Rutter2007,Mallet2007,Yang2010,Tapaszto2012,Xue2012,Koepke2013} in the study of graphene. 
On the other hand, transport properties are most commonly measured by using invasive macroscopic contacts. Such contacts represent only a minor perturbation in large systems, but can be the main source of scattering in nanoscale devices.
Here we evaluate the conductance between two STM-like tips, \ie , a situation where nanoscale transport properties can be extracted with noninvasive probes. The considered regime is thus between the single STM setup and the fixed macroscopic contacts.

The envisaged technique requires independently positioned point probes to act as input and output. Such setups have been achieved experimentally \cite{Hasegawa2002,Kubo2006,Jaschinsky2006,Kim2007,Baringhaus2013apl,Roychowdhury2014} and the recent progress is reviewed in detail in Refs. \onlinecite{Nakayama2012,Li2013}.
State-of-the-art experimental techniques \cite{Kubo2006,Baringhaus2014} allow for tip separations down to 50-100 nm.
Multi-probe measurements have been used to characterize several systems: anisotropic transport; \cite{Kanagawa2003} nanowires; \cite{Cherepanov2012,Kubo2006} carbon nanotubes; \cite{Watanabe2001} graphene nanoribbons;\cite{Baringhaus2014} grain boundaries both in graphene \cite{Clark2013,Clark2014} and other materials; \cite{Kim2010} and monolayer and bilayer graphene. \cite{Eder2013,Ji2012,Sutter2008}

 \begin{figure}[tb]
 \centering
 \includegraphics[width=0.8\columnwidth]
 {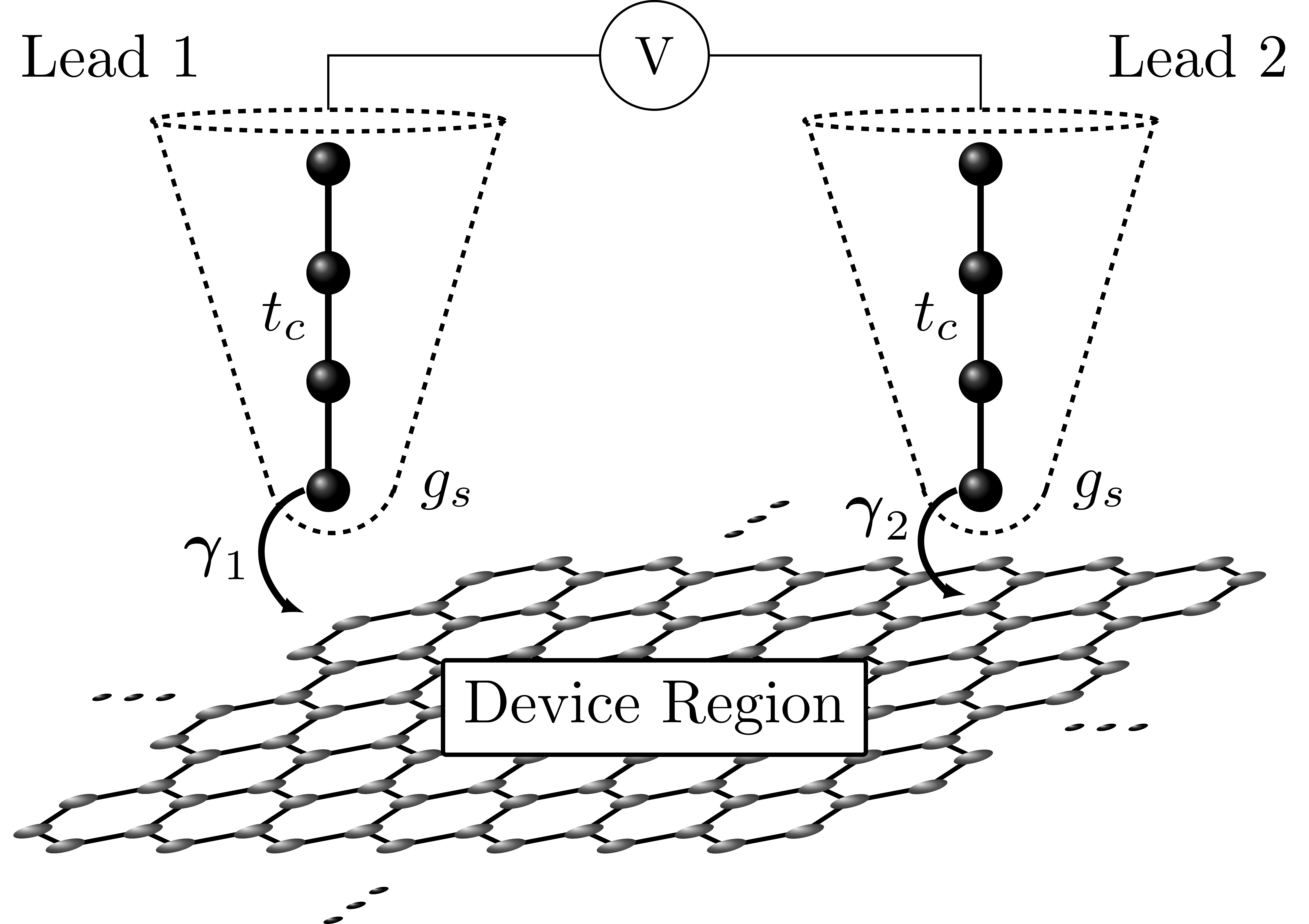}

 \caption[]{\small  Setup sketch including the leads modelled as one dimensional chains with a hopping $t_c$ between the sites. The surface Green's functions $g_s$ are indicated together with the coupling $\boldsymbol{\gamma}^{1/2}$ between the lead and the graphene sample.
  } \label{fig:setup}
 \end{figure}

Graphene-based materials are particularly interesting because the mean free path in high quality samples is comparable to or perhaps even longer than the probe separation.\cite{Baringhaus2014}
As a result, the dual-STM setup is effectively in the phase coherent regime at low temperatures. In this regime, structural details, such as single-site scattering centers, edges, or grain boundaries, limit the conductance, such that quantum interference phenomena become visible in the transmission between the probes.

In a recent work, \cite{Settnes2014} the present authors proposed a dual-probe setup on graphene with one fixed probe while the other probe operating in the scanning mode. We used real-space conductance maps to explore quantum
interference effects near defects and edges in graphene.
Fourier transforms of the real-space conductance maps allowed us to extract further details, and in particular they revealed information about intra and intervalley scattering due to these defects.
In the present work, we extend the theoretical investigation to the spectroscopic mode of the dual-probe system, where two fixed probes operate in the presence of an applied gate, which allows the Fermi energy to be varied.
While we focus on graphene as an illustrative example, particularly suited for the observation of quantum interference phenomena, the methodology is general and applicable to other surfaces or two-dimensional materials.
We use a combination of numerical calculations and analytic expressions to explain the spectroscopic fingerprints observed both in pristine graphene and in the presence of vacancies and adatoms.
Finally, we extend the framework to nanostructures such as perforations.

The paper is organized as follows: \secref{method} introduces the real space Green's function (GF) method and highlights the conceptual differences compared to standard recursive GF techniques.
\secref{Pristine} considers dual-probe spectroscopy of pristine graphene based on both analytical approximations of the GF and numerical calculations.
\secref{SimpleDefect} introduces defects where both single vacancies and adatoms are considered in the high symmetry directions, but also randomly placed in the sample.
Finally, \secref{AntiDot} considers perforations of the graphene lattice with different edge geometries.

\section{Method}\seclab{method}

\subsection{Transport calculations using point probes}
The transport setup consists of a device region and two leads as illustrated on \figref{setup}.
We describe the leads by the surface GF, $g_{S}$, which couples to the device region that is described by the retarded/advanced GF $\GGG/\GGG{}^\dagger$. We view $g_S$ as a known quantity (a simple analytic model is used below, but more elaborate models are readily incorporated in the formalism), and solve $\GGG/\GGG{}^\dagger$ from the appropriate Dyson equation, see below.
The main difference between the setup sketched in \figref{setup} and the standard Landauer setup \cite{DattaBook}, where left and right lead couple to the edge of a finite device region, is that the device region is now \textit{infinite}.
Standard recursive methods, treating infinite systems, use periodic boundary conditions. However, imposing periodic boundary conditions for the two-point probe setup would lead to a spurious repetition of the probes.
As a consequence we require a real space formalism ensuring that the probes only appear locally.

\subsection{Real space graphene Green's function}
The basic building block of our method is the real space representation of the GF for an infinite pristine graphene sheet, $\GGG{}^0$.  This object is computed using a nearest neighbour tight-binding model, and the  GF element connecting sites $i$ and $j$ is given by
\bal
G^0_{ij}(z) = \frac{1}{\Omega_{BZ}}\int\mathrm{d}^2\kkk \frac{N_{ij}(z)\ee^{\ii \kkk\cdot (\rrr_j-\rrr_i)}}{z^2 - \gamma_{cc}^2|f(\kkk)|^2} 
\eqlab{g0}
\eal
where $z = E +\ii 0^+$ is the energy, $\Omega_{BZ}$ is the area of the first Brillouin zone and the carbon-carbon hopping integral is $\gamma_{cc}\approx -2.7$ eV. \cite{Reich2002} The position of site $i$ is denoted by $\rrr_{i} = m_{i}\aaa_1+n_{i} \aaa_2$ in units of the lattice vectors $\aaa_1$ and $\aaa_2$ with $m_i$ and $n_i$ being integers. We introduced the definition $N_{ij}(z) = z$, when $i$ and $j$ are on the same sublattice and $N_{ij}(z) = \gamma_{cc}f(\kkk)$ if $i$ and $j$ are on opposite sublattices ($N_{ji} = N_{ij}^*$). Finally, we use the definition $f(\kkk) = 1 + \ee^{\ii \kkk \cdot\aaa_1} + \ee^{\ii \kkk \cdot\aaa_2}$.

The two dimensional integral \eqref{g0} can be converted to a single integration using complex contour techniques. \cite{Power2011} The remaining integration can be treated using standard numerical integration. This approach is valid regardless of the separation between the points $i$ and $j$; even long distance terms are easily obtained with only a minor complications arising in the convergence of \eqref{g0} due to the rapidly oscillating phase.

\subsection{Including defects} \seclab{Dyson}
To include modifications to the pristine lattice, we use the Dyson equation:
\bal
\GGG = \GGG{}^0 + \GGG{}^0 \VVV \GGG =  \big(\mathbf{1}-\GGG{}^0 \VVV\big)^{-1}\GGG{}^0, \eqlab{DysonEqn}
\eal
where $\GGG{}^0$ is the pristine GF matrix and $\VVV$ is the perturbation.
Any local perturbation (\eg, adatoms, vacancies, coupling to leads) can be included using this technique.
Accurate parametrizations for many of these perturbations can be obtained using density functional theory. \cite{Lherbier2012,RocheBook}

We note that the dimension of the $\VVV$-matrix is determined by the number of the modified sites.
Thus, for $N$  modified sites one needs to solve a $N\times N$ system, and the computational cost thus follows the number of defect and contact sites, rather than the sample size which is usual for recursive GF methods.

All perturbations to the pristine lattice are added in real space using \eqsref{g0}{DysonEqn}, as opposed to describing them with additional terms in the reciprocal space Hamiltonians.
This ensures that modifications are added locally and not repeated via periodic boundary conditions.
The approach is well suited to situations where the majority of the sample is pristine, as unmodified graphene is computationally ``free".

\subsection{Transmission}
The transmission coefficient between the two probes, $\calT_{12}$, yields the zero temperature conductance as $\mathcal{G}_{12} = \fracsmall{2e^2}{h}\calT_{12}$ (here we treat the spin degenerate case).
The transmission is given by \cite{AnttiBook,DattaBook}
\bal
\calT_{12}(E) = \mathrm{Tr}\big[\GGG(E)\mathbf{\Gamma}_1(E) \GGG {}^\dagger(E) \mathbf{\Gamma}_2(E)\big], \eqlab{TransEqn}
\eal
where $E$ is the energy, $\GGG$ is the full Green function \eqref{DysonEqn} (including the sites coupling to the leads) and $\mathbf{\Gamma}_{1/2}$ is the coupling to the leads given as $\mathbf{\Gamma}_{1/2}(E) = \ii\big(\mathbf{\Sigma}_{1/2} -\mathbf{\Sigma}_{1/2}^\dagger\big)$.
The self-energies $\mathbf{\Sigma}_{1/2}$ of the leads are calculated from the coupling matrix between the lead and the sample $\boldsymbol{\gamma}_{1/2}$ and the surface GF of the lead $g_{s}$, \ie \; $\mathbf{\Sigma}_{1/2} = \boldsymbol{\gamma}_{1/2}{}^\dagger g_{s} \boldsymbol{\gamma}_{1/2}$. We use a linear atomic chain model for the leads where the surface GF is known exactly:\cite{ecoBook} $g_S = \frac{E\pm \sqrt{E^2-4\gamma_{l}^2}}{2\gamma_{l}^2}$, where $\gamma_{l}$ is the coupling between the sites in the linear chain (here $\gamma_{l}=\gamma_{cc}$ is used). The parameters are chosen to ensure a constant DOS in the leads in the considered energy interval.

The coupling between the graphene and the tip of the probes is calculated using the Tersoff-Hamann approach \cite{Tersoff1983,Meunier1998}
\bal
\gamma_j = \gamma_0 w_j \ee^{-d_j/\lambda}\cos\big(\theta_j\big),
\eal
where $\theta_j$ and $d_j$ are the angle and the distance, respectively, between the tip apex and site $j$, $w_j = \ee^{-ad_j^2}/\sum_{m} \ee^{-ad_m^2}$, $\lambda=0.85 \textup{\AA}$ and $a=0.6 \textup{\AA}^{-2}$ are constants chosen in accordance to Refs.  \onlinecite{Meunier1998,Amara2007}.
$\gamma_0$ is a scaling factor, which in practical calculations is set to $\gamma_0 = 10\gamma_{cc}$.

When considering a probe coupling to a single site, the transmission in \eqref{TransEqn} reduces to the following simple form:
\bal
\calT_{12}(E) = \big(2\pi \gamma_1  \gamma_2  \rho_{lead}\big)^2 \; |G_{12}(E)|^2, \eqlab{TransSS}
\eal
where $\rho_{lead} = -\mathrm{Im}(g_{s})/\pi$ is the constant density of states of the last atom of lead.
Hence the only energy dependence originates from the GF term.

From \eqref{TransSS} we notice that the transmission scales with the DOS of the leads.
The transmission also scales with the coupling to the probes as $\sim \gamma_{1}^2 \gamma_2^2$.
As $\gamma_{1/2}$ depend exponentially on the distance between the tip and the sample, this means that decreasing the distance between sample and tip by $\sim 1 \; \textup{\AA}$ increases the coupling which in turn increases the transmission by a factor of $\sim 100$.

In what follows, we consider STM-like probes (\ie\; probes which couple only to a very limited number of sites in the sample) in order to obtain transparent results giving insight into the processes which dominate the transport between the point probes. More realistic or larger probes may be included within the presented framework by increasing the number of graphene lattice sites that couple to the probes or by substituting the semi-infinite mono-atomic chain by other surface GF's.

Finally, it is noted that we consider the low temperature and low bias regime and therefore ignore inelastic effects such as phonon scattering. \cite{Zhang2008} Here we also neglect the possible non-planarity of the graphene sheet, either due to the intrinsic ripples \cite{Fasolino2007} or  caused by one of the probes. \cite{Eder2013} However, we previously discussed the effect of ripples on dual-probe scanning mode calculations in Ref. \onlinecite{Settnes2014}.

\section{Pristine Graphene} \seclab{Pristine}

We first consider the case of pristine graphene without defects.
In this case we can gain a transparent understanding by the so called stationary phase approximation (SPA) \cite{Power2011} to the GF in \eqref{g0}.
The SPA is valid for the high symmetry directions (armchair or zigzag) and for separations between the $i$ and $j$ sites exceeding a few lattice spacings.
Using the SPA, the graphene GF in \eqref{g0} can be expressed as
\bsub \eqlab{SPA}
\bal
G_{ij,{\rm SPA}}^{0,ac} &= \frac{\calA(E) \ee^{\ii \calQ(E) d_{ij}}}{\sqrt{d_{ij}}} ,\eqlab{SPAa}\\
G_{ij,{\rm SPA}}^{0,zz} &= \sum_{\eta=\pm}\frac{\calA^\eta(E) \ee^{\ii \calQ^\eta(E) d_{ij}}}{\sqrt{d_{ij}}},\eqlab{SPAb}
\eal
\esub
where $\calA(E)$ is an energy dependent amplitude and $\calQ(E)$ is the Fermi wavevector in the armchair and zigzag directions.
The coefficients are given in \appref{SPAcoef} and derived in Ref. \onlinecite{Power2011}.

Inserting \eqref{SPA} into \eqref{TransSS} gives the distance dependence of the transmission, $\calT_{12} \propto 1/d_{12}$.
Consequently the resistance scales linearly with probe separation, $R\propto d_{12}$.

Consider now the case when the separation between the two probes is in the armchair direction. Using \eqref{SPAa}, we find that the transmission coefficient increases linearly with energy. The linear increase of $\calT^{(ac)} \propto |\calA|^2/d_{12}$ originates from the fact that $|\calA|^2$ grows linearly with energy for low energies, see \appref{SPAcoef}.

The zigzag direction is more complicated because of the two terms in \eqref{SPAb}, caused by the two non-identical sides of the Fermi surface along the zigzag direction:
\bal
\calT^{(zz)}_{12} &\times d_{12} \propto \big|\sum_{\eta=\pm} \calA^\eta \ee^{\ii \calQ^\eta d_{12}} \big|^2 \nonumber \\
&=|\calA^+|^2 + |\calA^-|^2 \nonumber \\
&\quad +|\calA^+||\calA^{-}|\cos\bigg(\bigg[\calQ^+-\calQ^-\bigg]d_{12}\bigg). \eqlab{SPAzz}
\eal
In addition to the linear increase (the first two terms), we also find an oscillating term.
The oscillation period decreases with increasing energy due to the energy dependence of $\calQ^+-\calQ^-$.
We therefore expect a more rapid oscillation for higher values of the Fermi energy.

 \begin{figure}[tb]
 \centering
 \begin{minipage}[b]{1\columnwidth}
\centering (a)
 \includegraphics[width=0.55\columnwidth]
 {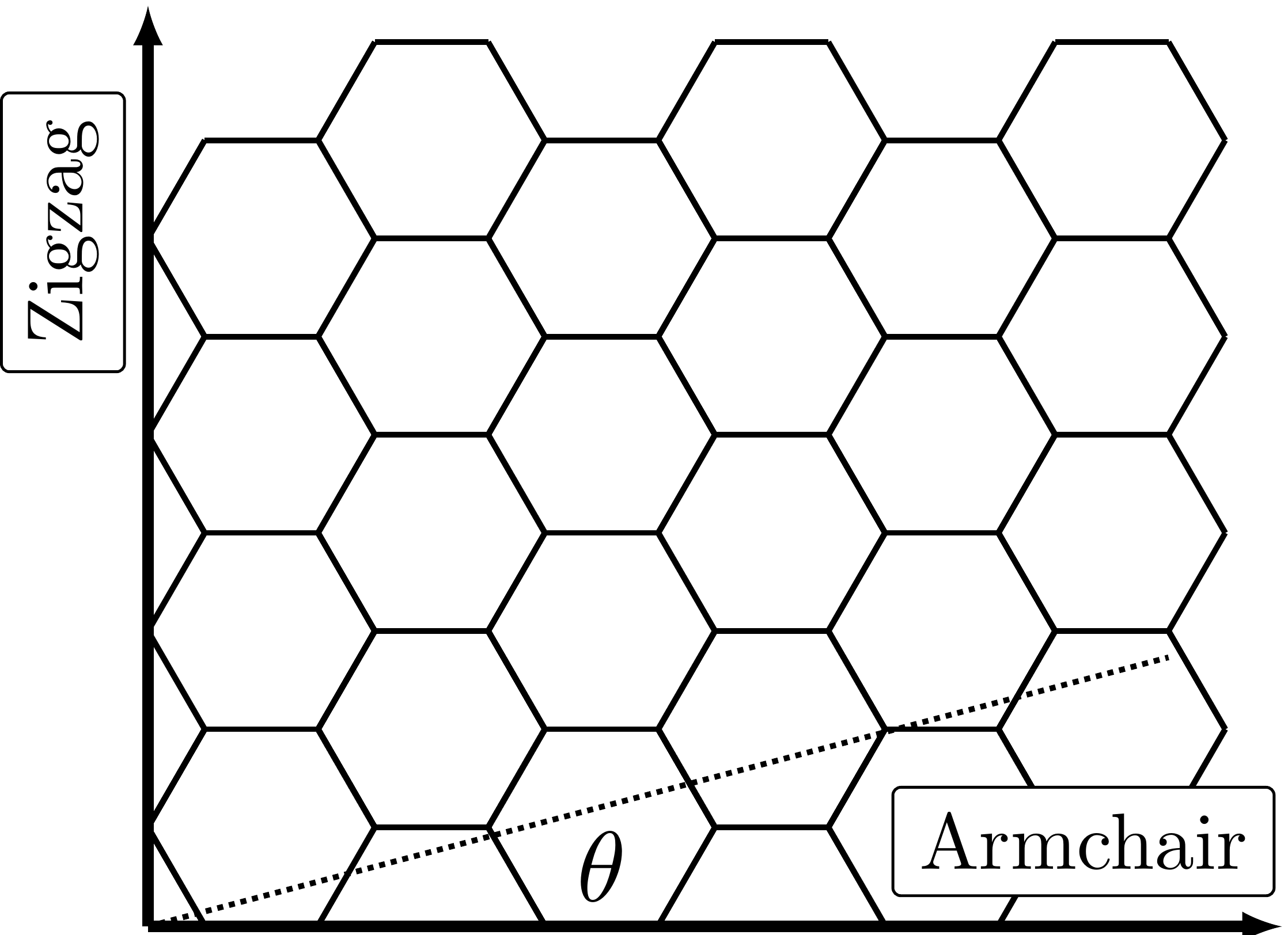}
 \end{minipage}

\vspace{0.5cm}
 \begin{minipage}[b]{1\columnwidth}
 \includegraphics[width=0.99\columnwidth]
 {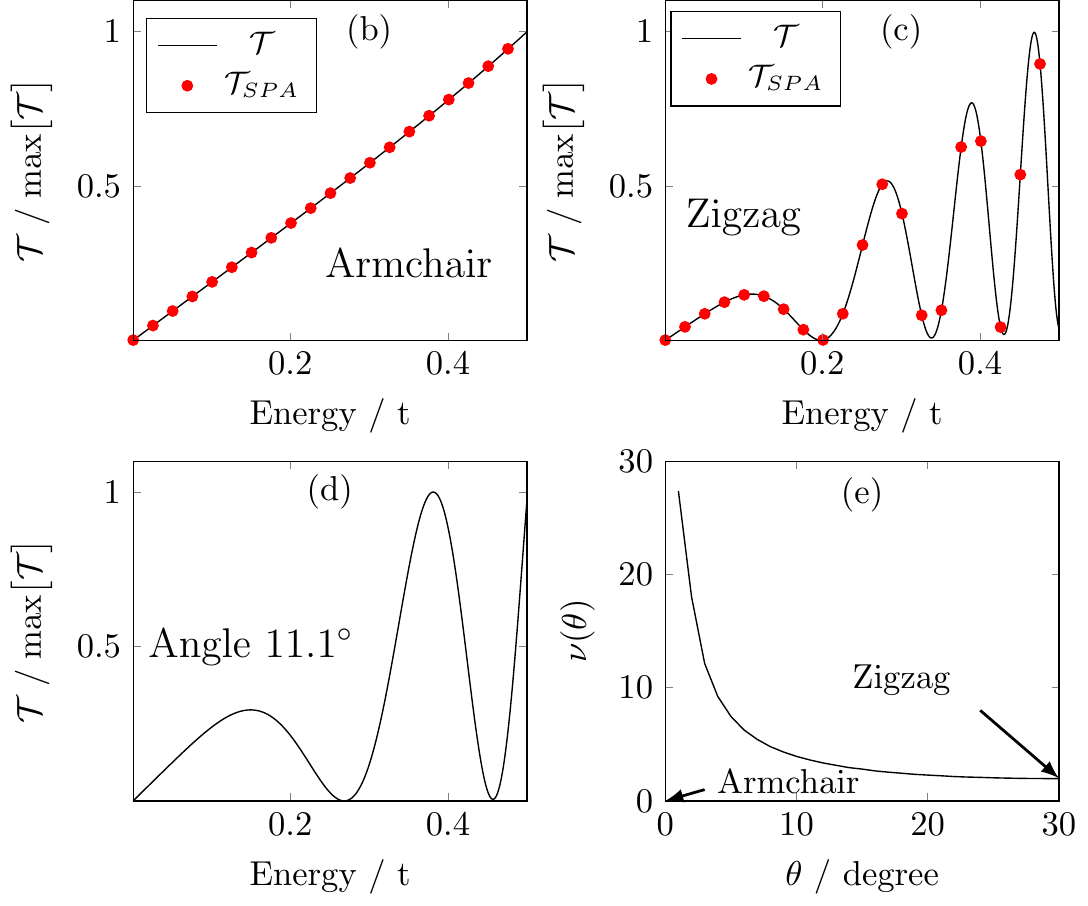}
 \end{minipage}

 \caption[]{\small (a) Sketch showing the pristine sample and the rotation angle $\theta$ from the armchair direction. (b-d) The transmission as a function of energy between the two leads separated by 50 nm along (b) armchair, (c) zigzag and (d) rotated $ \theta = 11.1^\circ$ from the armchair direction. In (b) and (c) the transmission calculated using the SPA is indicated (red dots). (e) The oscillation period $\nu(\theta)$ (see main text for definition) is plotted against rotation angle $\theta$ as defined in (a). The curve is constructed by averaging over many individual calculations with distances ranging from 20 to 100 nm.
  } \label{fig:sts_4panel}
 \end{figure}
 
In \figref{sts_4panel}b-c we plot the energy dependent transmission for $d_{ij}$ parallel to either armchair (b) and zigzag (c) for probe separation $\sim 50$ nm. The transmissions are calculated using both \eqref{SPA} (dots) and using a numerical evaluation of \eqref{g0} (line). We note an almost perfect match for all energies, which confirms the validity of the SPA approach.

In \figref{sts_4panel}d we consider a direction rotated $\theta \approx 11^\circ$ relative to the armchair direction. Consequently the oscillation period depends on the rotation angle $\theta$, as defined on \figref{sts_4panel}a.
The oscillation is a consequence of the asymmetry of the Fermi surface in the given direction and is therefore a fingerprint of the crystalline direction between the probes.

The GFs for all other separations (except armchair) have the same form as \eqref{SPAb}.\cite{Power2011} So the transmission generally takes a form equivalent to \eqref{SPAzz} but with different expressions for $\calQ^+$ and $\calQ^-$, which depend on the direction of separation.
In the limit of low energies we can expand the coefficients as $|\calA|^2 \propto E$ and $\big(\calQ^+-\calQ^-\big) \propto E^2 / \nu(\theta) $.
Here $\nu(\theta)$ is an oscillation period that depends on the angle $\theta$ (defined in \figref{sts_4panel}a).
Accordingly $\theta = 0$ denotes armchair separation and $\theta = 30^\circ$ denotes zigzag separation between the probes.
The energy dependence of the transmission in \eqref{SPAzz} now becomes
\bal
\calT_{12} \propto E \cos\big( E^2 d_{12} / \nu(\theta) + \phi_{ph}\big),
\eal
where $\phi_{ph}$ is a phase factor, which is independent of the direction but depends on the distance and the exact atoms coupling to the probes.
If we plot $\calT_{12}/E$ as a function of $d_{12} E^2$ we can determine the period $\nu(\theta)$ as the lowest full period of oscillation in the $\calT_{12}/E$ vs $d_{12} E^2$ plot for the corresponding angle $\theta$.
In \figref{sts_4panel}e, we plot $\nu(\theta)$ as a function of angle.  \figref{sts_4panel}e is the average of many individual calculations of $\nu(\theta)$ for separations ranging from 20 to 100 nm.

From \figref{sts_4panel}e we conclude that $\nu(\theta)$ provides a fingerprint of the probe separation direction. Furthermore $\nu(\theta)$ enables us to determine the crystalline direction with a simple spectroscopic measurement provided we know the distance between the probes and that the gate is kept sufficiently small.

\section{Simple Defects}\seclab{SimpleDefect}
Next we consider defects like vacancies and adatoms.
To obtain an analytical treatment in this case let the defects be coupled to a group of sites denoted $0$ and the probes coupled to sites denoted $1$ and $2$. We restate the Dyson equation (\eqref{DysonEqn}) using the $t$-matrix formalism
\bal
\GGG_{12} &= \GGG{}^0_{12} + \GGG{}^0_{10} {\mathbf{t}}_{00} \GGG{}^0_{02},
\eal
where
\bal
{\mathbf{t}}_{00} &=\big(\mathbf{1}-\VVV_{00}\GGG{}^0_{00}\big)^{-1}  \VVV_{00}.
\eal
Inserting this into \eqref{TransEqn} we obtain
\bal
\calT_{12} &\propto \mathrm{Tr}\big[ \big( \GGG_{12} + \GGG_{10}{\mathbf{t}}_{00}\GGG_{02}\big)\big(\GGG{}^{\dagger}_{12} + \GGG{}^{\dagger}_{02}{\mathbf{t}}_{00}^{\dagger}\GGG{}^{\dagger}_{10}\big)\big] \nonumber \\
&= \mathrm{Tr}\bigg[\GGG_{12}\GGG{}^\dagger_{12} +\big(\GGG_{10}{\mathbf{t}}_{00} \GGG_{02}\big)\big(\GGG_{10}{\mathbf{t}}_{00} \GGG_{02}\big)^\dagger \nonumber \\
&\quad + 2\mathrm{Re}\bigg\{(\GGG_{10}{\mathbf{t}}_{00}\GGG_{02})\GGG{}^{\dagger}_{12}\bigg\} \bigg]. \eqlab{TransDefect}
\eal
\eqref{TransDefect} is generally applicable. If the probes and the defect couple to single sites all matrices reduce to scalar quantities and enable simple analytic expressions. For example, we use the SPA expression \eqref{SPAa} when both probes and defects are along the armchair direction 
\bal
\Delta \calT_{12}& = \calT_{12} -\calT_{12}^0  \propto  \frac{|\calA|^4}{d_{10}d_{20}} |{t}_{00}|^2 \nonumber \\
&\quad - \frac{ |\calA |^3}{\sqrt{d_{10}d_{20}d_{12}}} \mathrm{Re}\bigg\{ (1+\ii){t}_{00} \ee^{\ii \calQ (d_{10}+d_{20}-d_{12})}\bigg\}, \eqlab{dTexp}
\eal
where $\calT_{12}^0$ is the pristine transmission, $d_{12}$ denotes the distance between the two probes, and $d_{10}$ and $d_{20}$ denote the distance between the defect site and probe 1 and 2, respectively.

Assuming the defect lies between the probes, \ie \; $d_{12} = d_{10}+d_{20}$, we get from \eqref{dTexp}
\bal
\Delta \calT_{12} & \propto   \frac{|\calA|^4}{d_{10}d_{20}} |{t}_{00}|^2- \frac{ |\calA |^3}{\sqrt{d_{10}d_{20}d_{12}}} \mathrm{Re}\bigg\{ (1+\ii){t}_{00}\bigg\}. \eqlab{TacSPA1}
\eal
Thus, a change in transmission occurs due to the backscattering at the defect.
This was also observed in Ref. \onlinecite{Settnes2014} where one probe scanned around the defect position to obtain a real space image of the transmission change.
The size and form of $\Delta \calT_{12}$ depend on the type of defect through ${t}_{00}$.

For the defect on either side of the probes, \ie \; $d_{10} = d_{12}+d_{20}$, \eqref{dTexp} becomes
\bal
\Delta \calT_{12} & \propto  \frac{|\calA|^4}{d_{10}d_{20}} |{t}_{00}|^2 - \frac{ |\calA |^3}{\sqrt{d_{10}d_{20}d_{12}}} \mathrm{Re}\bigg\{ (1+\ii){t}_{00}\ee^{2\ii \calQ d_{20}}\bigg\}. \eqlab{TacSPA2}
\eal
The result for the impurity on the other side of the probes ($d_{20} = d_{12}+d_{10}$) is obtained by interchanging $1$ and $2$.
The case in \eqref{TacSPA2} gives rise to oscillations as we change the energy (by changing $\calQ$).
The oscillations are a consequence of quantum interference between the outgoing wave from the output probe and the scattered wave.
Similar expressions as \eqsref{TacSPA1}{TacSPA2} can be derived for the zigzag separation, but the simple form is complicated by the two interfering terms in \eqref{SPAb}.

\eqsref{TacSPA1}{TacSPA2} show that the effect of the impurity enters through the $t$-matrix, which depends on the type of impurity.
In this section we consider two specific defects: vacancies and adatoms.
Vacancies are modelled as a change of the on-site energy, $V_{00}\rightarrow \infty$.
On the other hand, adatoms are modelled with an energy dependent self energy $\Sigma^\alpha$, describing a resonant level with energy $\epsilon_\alpha$, coupled to the graphene sample with coupling constant $\gamma_\alpha$, \ie \; $V_{00} = \Sigma^{\alpha}_{00}=|\gamma_\alpha|^2 / (E+\ii 0^+-\eps_\alpha)$.
The $t$-matrices become, \cite{Robinson2008,Wehling2010}
\bsub \eqlab{tmatrix}
\bal
\mathrm{Vacancy:}\;\; {t}_{00} &= \frac{V_{00}}{1-V_{00}G^0_{00}} \rightarrow - \frac{1}{G^0_{00}}.\\
\mathrm{Adatom:}\;\; {t}_{00} &= \frac{\Sigma^\alpha_{00}}{1-\Sigma^\alpha_{00}G^0_{00}} = \big(\Sigma^{\alpha}_{00}{}^{-1}-G^0_{00}\big)^{-1} \nonumber \\
&= \frac{|t_{\alpha}|^2}{E-\eps_\alpha -|t_\alpha|^2 G^0_{00}}.
\eal
\esub
The adatom gives rise to a resonant level whose position is determined by both $\eps_\alpha$ and $\gamma_\alpha$.
We choose parameters from Ref. \onlinecite{Uchoa2009} as $\eps_\alpha = -0.185|t|$ and $t_\alpha = 0.37 |t|$. This gives a resonant level within the energy interval of consideration.

 \begin{figure}[tb]
 \centering
 \includegraphics[width=0.90\columnwidth]
 {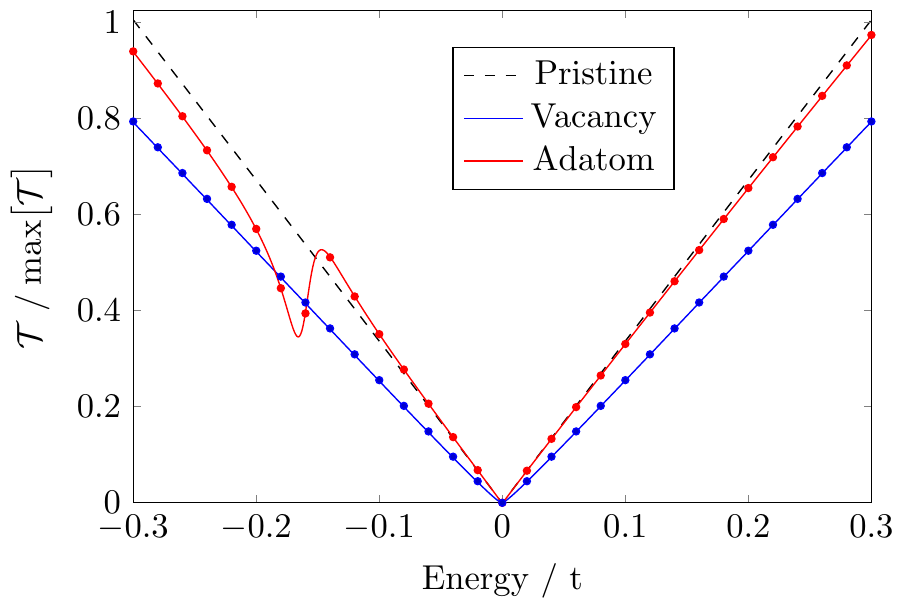}

 \caption[]{\small The transmission as a function of energy for pristine graphene (dashed), vacancy (red) and adatom (blue). The impurity is in between probes, which are separated by $\sim 50$ nm along the armchair direction. The dots denote a similar calculation using the SPA expression \eqref{TacSPA1}. The paramters for the adatom are chosen as in Ref. \onlinecite{Uchoa2009} as  $\eps_\alpha = -0.185|t|$ and $t_\alpha = 0.37 |t|$.
  } \label{fig:sts_defect}
 \end{figure}
 
\figref{sts_defect} shows the numerical result compared to the analytical expression \eqref{TacSPA1} for both a vacancy and an adatom. The impurities are located equidistant ($d_{10}=d_{20}=d_{12}/2$) from the two probes.
Again, we observe an almost perfect match between the analytic (symbols) and numerical (lines) results.
The vacancy gives rise to an overall reduction in transmission due to scattering, while the adatom leads to a smaller reduction of transmission, except at the resonance.
Especially at resonance the level of the adatom interacts strongly with the continuum of the graphene states giving rise to the asymmetric Fano type resonance \cite{Fano1961} observed at approximately $-0.15$ eV in \figref{sts_defect}.
Similar results are obtained for the zigzag direction, but superimposed onto the characteristic zigzag oscillation discussed in \secref{Pristine}.

 \begin{figure}[tb]
 \centering
 \begin{minipage}[tb]{1\columnwidth}
 (a) \includegraphics[width=0.7\columnwidth]
 {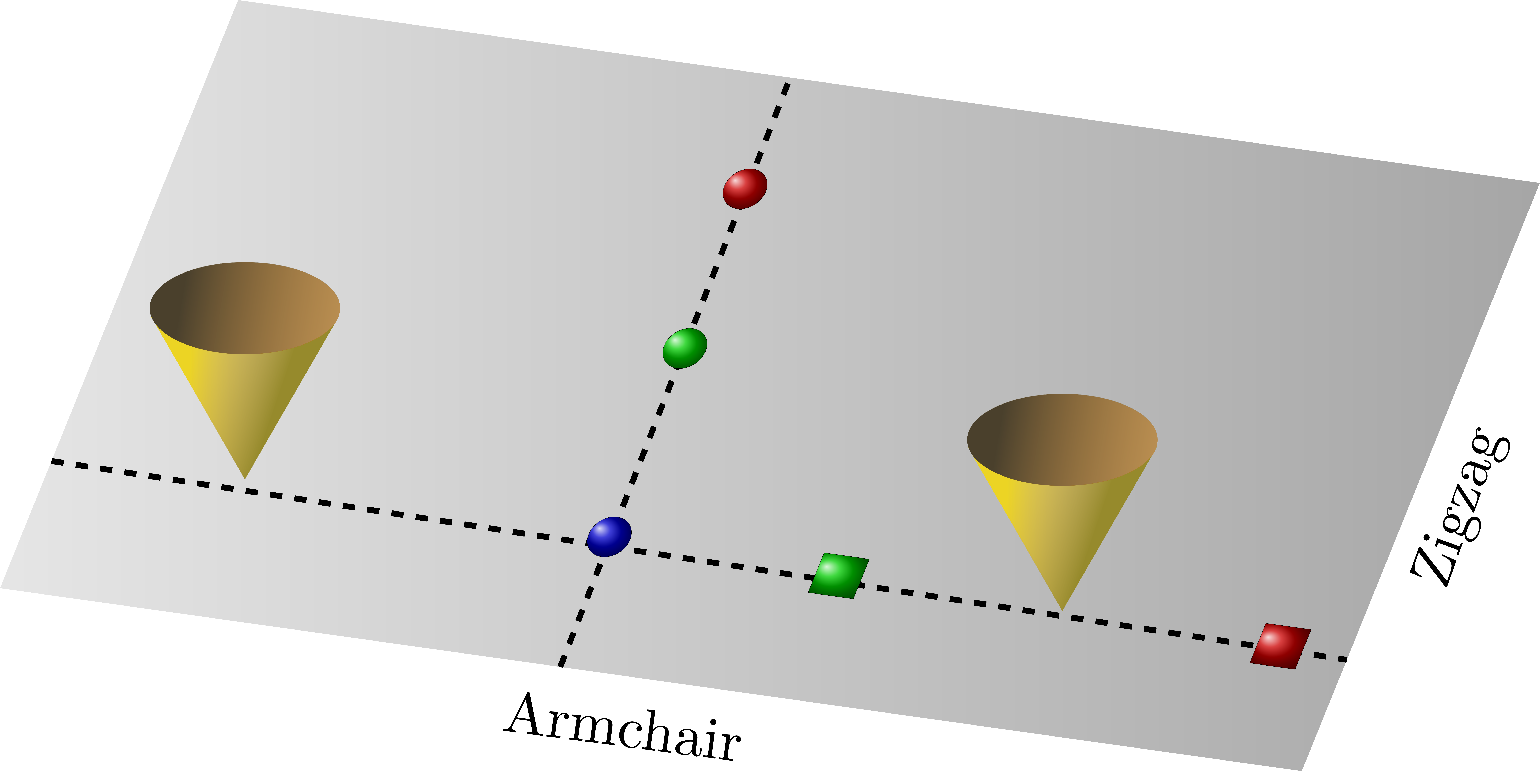} 
  \vspace{0.5cm}
 \end{minipage}

 \begin{minipage}[tb]{1\columnwidth}
 \centering
 \includegraphics[width=0.95\columnwidth]
 {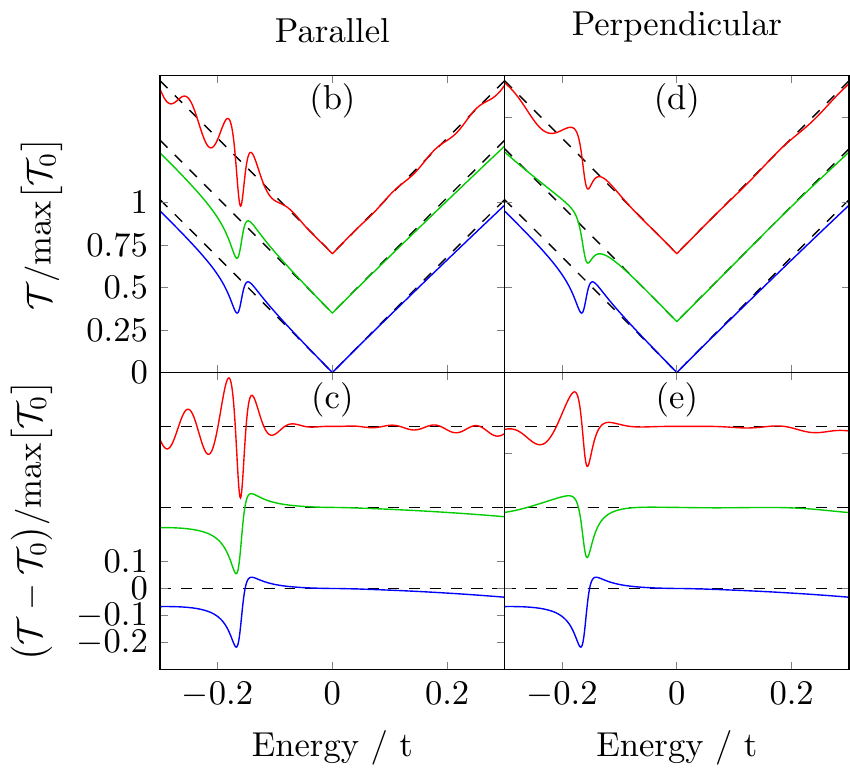}
 \end{minipage}
 \caption[]{\small (a) Sketch illustrating the two probes separated along the armchair direction by $\sim 50$ nm. The marks refers to impurity positions. Blue is along the line of separation and equidistant of the probes. The green and red squares are moved relative to the blue site along the armchair direction (parallel) by 12.8 nm and 34 nm, respectively. The transmission for the parallel translation are shown in (b) and (c).
The green and red circles are equidistant of the probes but moved along the zigzag direction (perpendicular) to 7.4 nm and 17.2 nm, respectively. The transmission function for impurities in these positions are shown in (d) and (e). The zero point for the curves has been translated for better distinction between curves.
  } \label{fig:sts_defect_panel}
 \end{figure}

 \begin{figure}[tb]
 \centering
 \includegraphics[width=0.8\columnwidth]
 {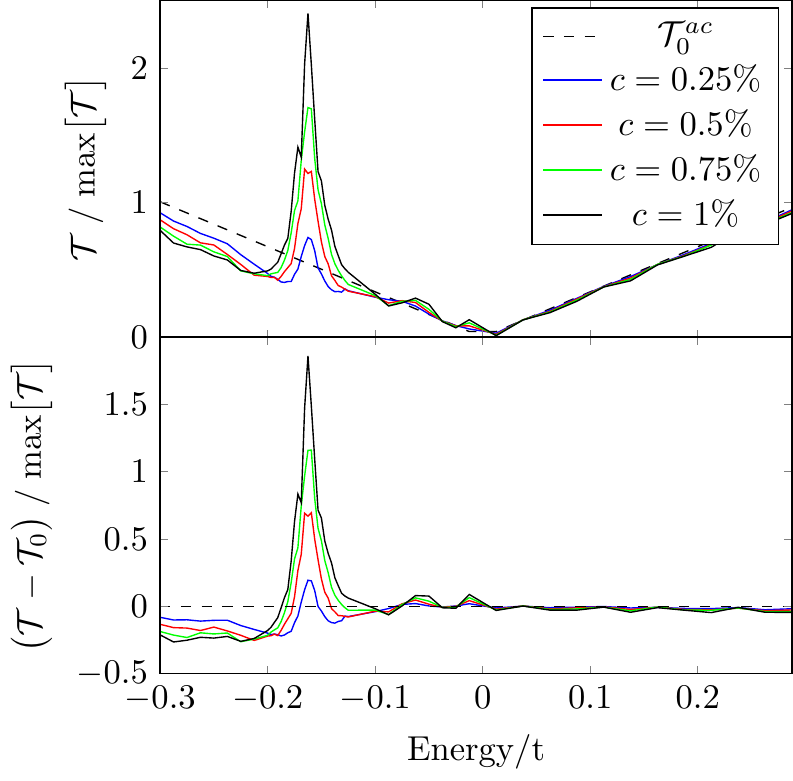}

 \caption[]{\small(a) Configuration averaged transmission as a function of energy. (b) The difference between the averaged transmission and the pristine transmission.
 We place impurities in a $50\times85$ nm square around the probes The unequal sides are chosen to take into account the probe separation direction. The curves are made from averaging $2\cdot 10^4$ configurations.
  } \label{fig:MCadatom}
 \end{figure}
 
 \subsection*{Impurity positions}
To investigate the influence of adatom position on the resonance, we now move the adatom away from the high symmetry point between the probes.
First, the adatom is moved along the line connecting the probes such that it is no longer equidistant from the probes.
These positions are shown by the red and green squares in \figref{sts_defect_panel}a. The corresponding dual-probe transmissions are shown in \figref{sts_defect_panel}b and the change relative to the pristine graphene sheet is shown in \figref{sts_defect_panel}c. Furthermore, both panels include the transmission for the equidistant impurity (blue curve) for comparison.
Likewise Figs. \ref{fig:sts_defect_panel}d and \ref{fig:sts_defect_panel}e show the corresponding transmissions as the adatom is moved perpendicular to the line separating the probes while keeping the impurity equidistant to the probes.

First, we consider the parallel case. Here the adatom is either in-between the probes, yet closer to one of them, or to the far side of one of the probes (\ie \; green and red square on \figref{sts_defect_panel}).
The Fano-type resonance persists as the adatom is moved. Only the form of the resonance changes.
However, we notice a distinct difference between the two cases.
When the impurity does not lie between the probes (red square), additional oscillations arise. This can be understood by comparing \eqsref{TacSPA1}{TacSPA2} where the difference is the term, $\exp\big[2i\calQ d_{20}\big]$. This term gives rise to oscillations through the energy dependence of $\calQ$. The oscillations have the same origin as these investigated in real space in Ref. \onlinecite{Settnes2014}, while scanning one probe around the impurity.
We notice the same effect for vacancy positions everywhere outside the high symmetry positions.

The same type of oscillations are present for the perpendicular direction. In this case we have to consider the interference between the emitted wave and the scattered wave returning from the impurity in the direction of the second probe.

\subsection*{Configurational average}
In an experimental setup, however, individual defects or adatoms can be difficult to locate.
This makes investigations of many randomly scattered defects important.
We fix the two probes with an armchair separation of 50 nm and place adatoms randomly with varying concentration.
The averaged transmissions are shown in \figref{MCadatom}.
The transmission is almost unchanged at energies away from the resonance, despite the oscillations caused by individual impurity positions shown in \figref{sts_defect_panel}.
This shows that the oscillations, induced by interference between incoming and scattered waves, tend to average out for many defects.
However, the resonance feature persists through configurational averaging as is evident from \figref{MCadatom}.
The signal is enhanced on resonance and an overall Fano type resonance is present in \figref{MCadatom}b with a height that scales with impurity concentration.
This suggests that the dual-probe setup can detect the type (position of resonant level) and concentration (peak height) of adatoms on the surface of a graphene sample.
This is in line with the suggested applications of graphene as a gas sensor. \cite{Schedin2007,Cagliani2014}
In the case of random vacancies we see an overall decrease in the transmission following the impurity concentration.
In this case a zero energy peak is present due to localisation effects around vacancies. This feature has been described in several works addressing the LDOS \cite{Pellegrino2009,Peres2006,Lawlor2013}.

\section{Perforated graphene}\seclab{AntiDot}
Many applications require deliberate nanostructuring of the graphene in order to engineer its electronic structure.
Therefore tools to investigate the transport properties of individual nanostructures are important in order to confirm the fabricated structure and its influence on nanoscale electron transport.

\begin{figure}[tb]
 \centering
 \includegraphics[width=0.99\columnwidth]
 {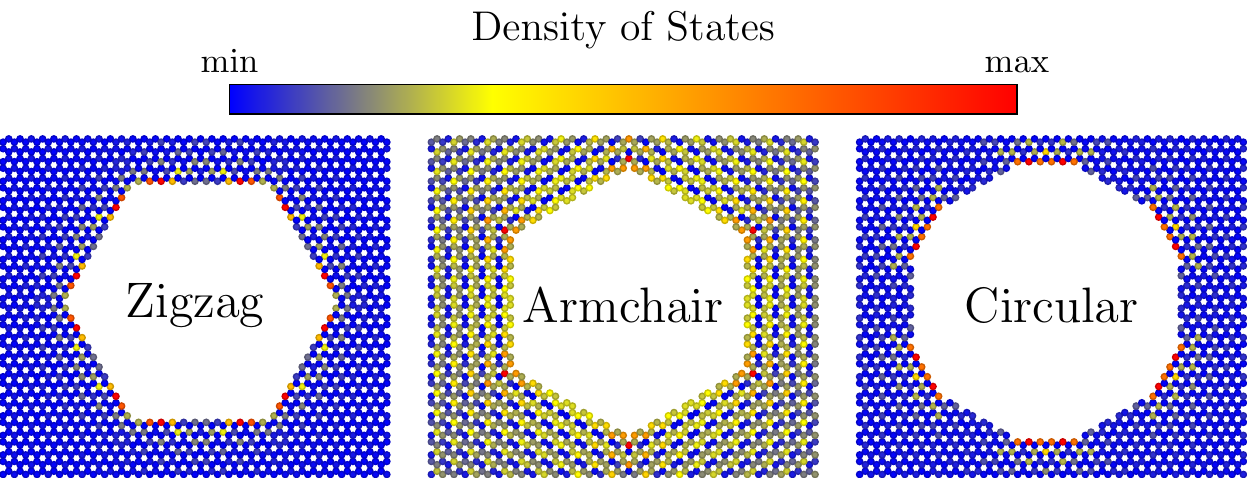}

 \caption[]{\small The density of states for  $E = 0.028|t|$ around antidots with different edge structures as indicated. The maps are individually scaled.
  } \label{fig:ldos_antidot_types}
 \end{figure}

In this section we consider perforations in the pristine graphene sheet -- so called antidots (see \figref{ldos_antidot_types}). \cite{Furst2009}
Several studies \cite{Pedersen2008,Pedersen2012,Gunst2011} show that arrays of antidots can induce a bandgap in graphene. 
The effect of antidots on the electronic properties of graphene strongly depends on the exact edge geometries of the antidots.
Therefore it is important to study the formation of single antidots and determine their edge configuration.

The perforations are modelled by removing the hopping matrix elements between sites around the edge of the hole, effectively disconnecting the sites from the rest of the graphene lattice.

 \begin{figure}[tb]
 \centering
 \includegraphics[width=0.99\columnwidth]
 {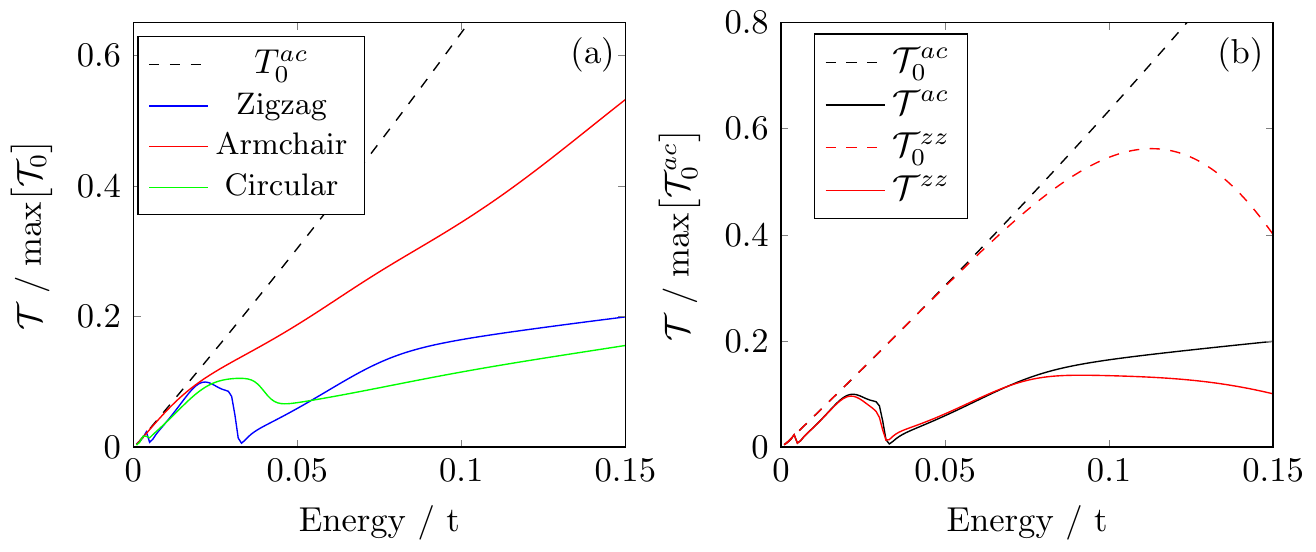}

 \caption[]{\small (a) The transmission for probes separated along the armchair direction ($\sim 50$ nm) for zigzag, armchair and circular antidots, respectively. The antidot structures are shown in \figref{ldos_antidot_types}. (b) The transmission for the same zig-zag antidot as (a), with probe separations ($\sim 50$ nm) along armchair and zig-zag direction, respectively.
  } \label{fig:sts_antidot12_panel}
 \end{figure}

We consider three possible edge geometries for antidots: zigzag, armchair or circular, the last contains an alternating sequence of armchair and zigzag edges (\figref{ldos_antidot_types}).
We calculate the transmission for each antidot type placed between probes separated in the armchair direction. The result is shown in \figref{sts_antidot12_panel}. As expected the transmission is generally lowered by introduction of the perforation.
A notable difference between the antidot types is a transmission dip present for the circular and especially zigzag type antidots.
This dip resembles the Fano type resonance observed for single adatoms, \figref{sts_defect}.
\figref{sts_antidot12_panel}a suggests that the resonant feature is connected to the zigzag edges, as the circular antidot consists of a mixture of zigzag and armchair edges. We therefore map the local density of states on sites around the antidot at the energy of the transmission dip (\cf\; \figref{ldos_antidot_types}).
The DOS is localised around the zigzag edges as discussed in Ref. \onlinecite{Gunst2011}.
These localised zigzag edge states being essentially dispersionless resemble a single level and therefore create a Fano type resonance in the transmission for antidots possessing zigzag edges. In addition,
we notice the difference between the resonance of the circular and zigzag antidot on \figref{sts_antidot12_panel}a.
The resonance of the pure zigzag edge has a sharper feature than the mixed edge (circular antidot). This leads to the conclusion that the resonance features can be related to the amount of zigzag edge present.
This in turn can be used as a fingerprint to determine the edge profile of antidots and other nanostructures.

A necessary condition for the dual-probe setup to be a useful tool for characterization of larger nanostructures, such as antidots, is that the exact direction between the probes should not have a great impact on the spectroscopic fingerprints of the nanostructure.
Therefore we compare the transmission for both zigzag and armchair probe separations in \figref{sts_antidot12_panel}b and note very similar behaviour at low energies.
This shows that especially the Fano-type resonance is not particularly sensitive to the orientation of the probes, and that the two-probe setup indeed can be a useful characterization tool for the electrical properties of individual nanstructures.

\section{Conclusion}
A dual-probe setup with probe separation distances in the nanometer range, makes it possible to obtain local transport properties on the nanoscale. We have presented a theoretical treatment of such a setup based on a real space Green's function method. 
This allows calculation of the transmission between two point probes on an infinite graphene sheet, without requiring periodicity of either probe or sample, while keeping the computational size proportional to the number of modified sites, as opposed to proportional to the total system size. 

Directional transport effects, not directly attainable using macroscopic contacts, are explored together with the spectroscopic fingerprints of local perturbations such as vacancies and adatoms. 
Additionally we show the capability of the dual-probe system to characterize nanostructures. In particular, we observe Fano-type resonances arising from resonant states in adatoms or near edges witha zigzag geometry.

The demonstrated features of the dual-probe setup, like conductance mapping\cite{Settnes2014} and spectroscopic analysis, suggest that it has a high potential for applications in the exploration of transport properties on the nanometer-scale.

\textbf{Acknowledgements}
The Center for Nanostructured Graphene (CNG) is sponsored by the Danish Research Foundation, Project DNRF58.
\appendix
\section{Coefficients for the SPA} \seclab{SPAcoef}
Below are given the coefficients for the SPA expressions \eqref{SPA} as derived in \onlinecite{Power2011}
\bsub
\bal
\calQ(E) &= \pm\cos^{-1} \bigg(-\sqrt{1-\fracsmall{E^2}{t^2}}\bigg)\\
\calA(E) &= -\frac{1+\ii}{\sqrt{\pi}} \frac{\sqrt{E}}{\sqrt{(E^2+3t^2)\sqrt{t^2-E^2}}}\\
\calQ^+(E) &= \pm\cos^{-1}\bigg(\frac{-t-E}{2t}\bigg)\\
\calQ^-(E) &=\pm\cos^{-1}\bigg(\frac{-t+E}{2t}\bigg) \\
\calA^+(E) &= -\frac{1+1i}{\sqrt{4\pi}} \sqrt{\frac{E}{|t|(t+E)}} \frac{1}{\big((3t+E)(t-E)\big)^{1/4}} \\
\calA^-(E) &= -\frac{1+1i}{\sqrt{4\pi}} \sqrt{\frac{E}{|t|(t-E)}} \frac{1}{\big((3t-E)(t+E)\big)^{1/4}}
\eal\eqlab{SPAcoef}
\esub


\begin{thebibliography}{63}%
\makeatletter
\providecommand \@ifxundefined [1]{%
 \@ifx{#1\undefined}
}%
\providecommand \@ifnum [1]{%
 \ifnum #1\expandafter \@firstoftwo
 \else \expandafter \@secondoftwo
 \fi
}%
\providecommand \@ifx [1]{%
 \ifx #1\expandafter \@firstoftwo
 \else \expandafter \@secondoftwo
 \fi
}%
\providecommand \natexlab [1]{#1}%
\providecommand \enquote  [1]{``#1''}%
\providecommand \bibnamefont  [1]{#1}%
\providecommand \bibfnamefont [1]{#1}%
\providecommand \citenamefont [1]{#1}%
\providecommand \href@noop [0]{\@secondoftwo}%
\providecommand \href [0]{\begingroup \@sanitize@url \@href}%
\providecommand \@href[1]{\@@startlink{#1}\@@href}%
\providecommand \@@href[1]{\endgroup#1\@@endlink}%
\providecommand \@sanitize@url [0]{\catcode `\\12\catcode `\$12\catcode
  `\&12\catcode `\#12\catcode `\^12\catcode `\_12\catcode `\%12\relax}%
\providecommand \@@startlink[1]{}%
\providecommand \@@endlink[0]{}%
\providecommand \url  [0]{\begingroup\@sanitize@url \@url }%
\providecommand \@url [1]{\endgroup\@href {#1}{\urlprefix }}%
\providecommand \urlprefix  [0]{URL }%
\providecommand \Eprint [0]{\href }%
\providecommand \doibase [0]{http://dx.doi.org/}%
\providecommand \selectlanguage [0]{\@gobble}%
\providecommand \bibinfo  [0]{\@secondoftwo}%
\providecommand \bibfield  [0]{\@secondoftwo}%
\providecommand \translation [1]{[#1]}%
\providecommand \BibitemOpen [0]{}%
\providecommand \bibitemStop [0]{}%
\providecommand \bibitemNoStop [0]{.\EOS\space}%
\providecommand \EOS [0]{\spacefactor3000\relax}%
\providecommand \BibitemShut  [1]{\csname bibitem#1\endcsname}%
\let\auto@bib@innerbib\@empty
\bibitem [{\citenamefont {Novoselov}\ \emph {et~al.}(2005)\citenamefont
  {Novoselov}, \citenamefont {Jiang}, \citenamefont {Schedin}, \citenamefont
  {Booth}, \citenamefont {Khotkevich}, \citenamefont {Morozov},\ and\
  \citenamefont {Geim}}]{Novoselov2005}%
  \BibitemOpen
  \bibfield  {author} {\bibinfo {author} {\bibfnamefont {K.~S.}\ \bibnamefont
  {Novoselov}}, \bibinfo {author} {\bibfnamefont {D.}~\bibnamefont {Jiang}},
  \bibinfo {author} {\bibfnamefont {F.}~\bibnamefont {Schedin}}, \bibinfo
  {author} {\bibfnamefont {T.~J.}\ \bibnamefont {Booth}}, \bibinfo {author}
  {\bibfnamefont {V.~V.}\ \bibnamefont {Khotkevich}}, \bibinfo {author}
  {\bibfnamefont {S.~V.}\ \bibnamefont {Morozov}}, \ and\ \bibinfo {author}
  {\bibfnamefont {a.~K.}\ \bibnamefont {Geim}},\ }\href {\doibase
  10.1073/pnas.0502848102} {\bibfield  {journal} {\bibinfo  {journal}
  {Proceedings of the National Academy of Sciences of the United States of
  America}\ }\textbf {\bibinfo {volume} {102}},\ \bibinfo {pages} {10451}
  (\bibinfo {year} {2005})}\BibitemShut {NoStop}%
\bibitem [{\citenamefont {Dean}\ \emph {et~al.}(2010)\citenamefont {Dean},
  \citenamefont {Young}, \citenamefont {Meric}, \citenamefont {Lee},
  \citenamefont {Wang}, \citenamefont {Sorgenfrei}, \citenamefont {Watanabe},
  \citenamefont {Taniguchi}, \citenamefont {Kim}, \citenamefont {Shepard},\
  and\ \citenamefont {Hone}}]{Dean2010}%
  \BibitemOpen
  \bibfield  {author} {\bibinfo {author} {\bibfnamefont {C.~R.}\ \bibnamefont
  {Dean}}, \bibinfo {author} {\bibfnamefont {A.~F.}\ \bibnamefont {Young}},
  \bibinfo {author} {\bibfnamefont {I.}~\bibnamefont {Meric}}, \bibinfo
  {author} {\bibfnamefont {C.}~\bibnamefont {Lee}}, \bibinfo {author}
  {\bibfnamefont {L.}~\bibnamefont {Wang}}, \bibinfo {author} {\bibfnamefont
  {S.}~\bibnamefont {Sorgenfrei}}, \bibinfo {author} {\bibfnamefont
  {K.}~\bibnamefont {Watanabe}}, \bibinfo {author} {\bibfnamefont
  {T.}~\bibnamefont {Taniguchi}}, \bibinfo {author} {\bibfnamefont
  {P.}~\bibnamefont {Kim}}, \bibinfo {author} {\bibfnamefont {K.~L.}\
  \bibnamefont {Shepard}}, \ and\ \bibinfo {author} {\bibfnamefont
  {J.}~\bibnamefont {Hone}},\ }\href {\doibase 10.1038/nnano.2010.172}
  {\bibfield  {journal} {\bibinfo  {journal} {Nature Nanotechnology}\ }\textbf
  {\bibinfo {volume} {5}},\ \bibinfo {pages} {722} (\bibinfo {year}
  {2010})}\BibitemShut {NoStop}%
\bibitem [{\citenamefont {Wang}\ \emph {et~al.}(2012)\citenamefont {Wang},
  \citenamefont {Kalantar-Zadeh}, \citenamefont {Kis}, \citenamefont
  {Coleman},\ and\ \citenamefont {Strano}}]{Wang2012}%
  \BibitemOpen
  \bibfield  {author} {\bibinfo {author} {\bibfnamefont {Q.~H.}\ \bibnamefont
  {Wang}}, \bibinfo {author} {\bibfnamefont {K.}~\bibnamefont
  {Kalantar-Zadeh}}, \bibinfo {author} {\bibfnamefont {A.}~\bibnamefont {Kis}},
  \bibinfo {author} {\bibfnamefont {J.~N.}\ \bibnamefont {Coleman}}, \ and\
  \bibinfo {author} {\bibfnamefont {M.~S.}\ \bibnamefont {Strano}},\ }\href
  {\doibase 10.1038/nnano.2012.193} {\bibfield  {journal} {\bibinfo  {journal}
  {Nature nanotechnology}\ }\textbf {\bibinfo {volume} {7}},\ \bibinfo {pages}
  {699} (\bibinfo {year} {2012})}\BibitemShut {NoStop}%
\bibitem [{\citenamefont {{Castro Neto}}\ \emph {et~al.}(2009)\citenamefont
  {{Castro Neto}}, \citenamefont {Peres}, \citenamefont {Novoselov},\ and\
  \citenamefont {Geim}}]{CastroNeto2009}%
  \BibitemOpen
  \bibfield  {author} {\bibinfo {author} {\bibfnamefont {A.~H.}\ \bibnamefont
  {{Castro Neto}}}, \bibinfo {author} {\bibfnamefont {N.~M.~R.}\ \bibnamefont
  {Peres}}, \bibinfo {author} {\bibfnamefont {K.~S.}\ \bibnamefont
  {Novoselov}}, \ and\ \bibinfo {author} {\bibfnamefont {A.~K.}\ \bibnamefont
  {Geim}},\ }\href {\doibase 10.1103/RevModPhys.81.109} {\bibfield  {journal}
  {\bibinfo  {journal} {Reviews of Modern Physics}\ }\textbf {\bibinfo {volume}
  {81}},\ \bibinfo {pages} {109} (\bibinfo {year} {2009})}\BibitemShut
  {NoStop}%
\bibitem [{\citenamefont {Peres}\ \emph {et~al.}(2006)\citenamefont {Peres},
  \citenamefont {Guinea},\ and\ \citenamefont {{Castro Neto}}}]{Peres2006}%
  \BibitemOpen
  \bibfield  {author} {\bibinfo {author} {\bibfnamefont {N.~M.~R.}\
  \bibnamefont {Peres}}, \bibinfo {author} {\bibfnamefont {F.}~\bibnamefont
  {Guinea}}, \ and\ \bibinfo {author} {\bibfnamefont {A.~H.}\ \bibnamefont
  {{Castro Neto}}},\ }\href {\doibase 10.1103/PhysRevB.73.125411} {\bibfield
  {journal} {\bibinfo  {journal} {Physical Review B}\ }\textbf {\bibinfo
  {volume} {73}},\ \bibinfo {pages} {125411} (\bibinfo {year}
  {2006})}\BibitemShut {NoStop}%
\bibitem [{\citenamefont {Chen}\ \emph {et~al.}(2008)\citenamefont {Chen},
  \citenamefont {Jang}, \citenamefont {Xiao}, \citenamefont {Ishigami},\ and\
  \citenamefont {Fuhrer}}]{Chen2008}%
  \BibitemOpen
  \bibfield  {author} {\bibinfo {author} {\bibfnamefont {J.-H.}\ \bibnamefont
  {Chen}}, \bibinfo {author} {\bibfnamefont {C.}~\bibnamefont {Jang}}, \bibinfo
  {author} {\bibfnamefont {S.}~\bibnamefont {Xiao}}, \bibinfo {author}
  {\bibfnamefont {M.}~\bibnamefont {Ishigami}}, \ and\ \bibinfo {author}
  {\bibfnamefont {M.~S.}\ \bibnamefont {Fuhrer}},\ }\href {\doibase
  10.1038/nnano.2008.58} {\bibfield  {journal} {\bibinfo  {journal} {Nature
  nanotechnology}\ }\textbf {\bibinfo {volume} {3}},\ \bibinfo {pages} {206}
  (\bibinfo {year} {2008})}\BibitemShut {NoStop}%
\bibitem [{\citenamefont {Binnig}\ \emph {et~al.}(1982)\citenamefont {Binnig},
  \citenamefont {Rohrer}, \citenamefont {Gerber},\ and\ \citenamefont
  {Weibel}}]{Binnig1982}%
  \BibitemOpen
  \bibfield  {author} {\bibinfo {author} {\bibfnamefont {G.}~\bibnamefont
  {Binnig}}, \bibinfo {author} {\bibfnamefont {H.}~\bibnamefont {Rohrer}},
  \bibinfo {author} {\bibfnamefont {C.}~\bibnamefont {Gerber}}, \ and\ \bibinfo
  {author} {\bibfnamefont {E.}~\bibnamefont {Weibel}},\ }\href {\doibase
  10.1103/PhysRevLett.49.57} {\bibfield  {journal} {\bibinfo  {journal} {Phys.
  Rev. Lett.}\ }\textbf {\bibinfo {volume} {49}},\ \bibinfo {pages} {57}
  (\bibinfo {year} {1982})}\BibitemShut {NoStop}%
\bibitem [{\citenamefont {Deshpande}\ and\ \citenamefont
  {LeRoy}(2012)}]{Deshpande2012}%
  \BibitemOpen
  \bibfield  {author} {\bibinfo {author} {\bibfnamefont {A.}~\bibnamefont
  {Deshpande}}\ and\ \bibinfo {author} {\bibfnamefont {B.~J.}\ \bibnamefont
  {LeRoy}},\ }\href {\doibase http://dx.doi.org/10.1016/j.physe.2011.11.024}
  {\bibfield  {journal} {\bibinfo  {journal} {Physica E: Low-dimensional
  Systems and Nanostructures}\ }\textbf {\bibinfo {volume} {44}},\ \bibinfo
  {pages} {743 } (\bibinfo {year} {2012})}\BibitemShut {NoStop}%
\bibitem [{\citenamefont {Cheianov}\ and\ \citenamefont
  {Fal'ko}(2006)}]{Cheianov2006}%
  \BibitemOpen
  \bibfield  {author} {\bibinfo {author} {\bibfnamefont {V.~V.}\ \bibnamefont
  {Cheianov}}\ and\ \bibinfo {author} {\bibfnamefont {V.~I.}\ \bibnamefont
  {Fal'ko}},\ }\href {\doibase 10.1103/PhysRevLett.97.226801} {\bibfield
  {journal} {\bibinfo  {journal} {Physical Review Letters}\ }\textbf {\bibinfo
  {volume} {97}},\ \bibinfo {pages} {226801} (\bibinfo {year}
  {2006})}\BibitemShut {NoStop}%
\bibitem [{\citenamefont {Bena}(2008)}]{Bena2008}%
  \BibitemOpen
  \bibfield  {author} {\bibinfo {author} {\bibfnamefont {C.}~\bibnamefont
  {Bena}},\ }\href {\doibase 10.1103/PhysRevLett.100.076601} {\bibfield
  {journal} {\bibinfo  {journal} {Physical Review Letters}\ }\textbf {\bibinfo
  {volume} {100}},\ \bibinfo {pages} {076601} (\bibinfo {year}
  {2008})}\BibitemShut {NoStop}%
\bibitem [{\citenamefont {Pellegrino}\ \emph {et~al.}(2009)\citenamefont
  {Pellegrino}, \citenamefont {Angilella},\ and\ \citenamefont
  {Pucci}}]{Pellegrino2009}%
  \BibitemOpen
  \bibfield  {author} {\bibinfo {author} {\bibfnamefont {F.~M.~D.}\
  \bibnamefont {Pellegrino}}, \bibinfo {author} {\bibfnamefont {G.~G.~N.}\
  \bibnamefont {Angilella}}, \ and\ \bibinfo {author} {\bibfnamefont
  {R.}~\bibnamefont {Pucci}},\ }\href {\doibase 10.1103/PhysRevB.80.094203}
  {\bibfield  {journal} {\bibinfo  {journal} {Physical Review B}\ }\textbf
  {\bibinfo {volume} {80}},\ \bibinfo {pages} {094203} (\bibinfo {year}
  {2009})}\BibitemShut {NoStop}%
\bibitem [{\citenamefont {Peres}\ \emph {et~al.}(2009)\citenamefont {Peres},
  \citenamefont {Yang},\ and\ \citenamefont {Tsai}}]{Peres2009}%
  \BibitemOpen
  \bibfield  {author} {\bibinfo {author} {\bibfnamefont {N.~M.~R.}\
  \bibnamefont {Peres}}, \bibinfo {author} {\bibfnamefont {L.}~\bibnamefont
  {Yang}}, \ and\ \bibinfo {author} {\bibfnamefont {S.-W.}\ \bibnamefont
  {Tsai}},\ }\href {\doibase 10.1088/1367-2630/11/9/095007} {\bibfield
  {journal} {\bibinfo  {journal} {New Journal of Physics}\ }\textbf {\bibinfo
  {volume} {11}},\ \bibinfo {pages} {095007} (\bibinfo {year}
  {2009})}\BibitemShut {NoStop}%
\bibitem [{\citenamefont {M\'{a}rk}\ \emph {et~al.}(2012)\citenamefont
  {M\'{a}rk}, \citenamefont {Vancs\'{o}}, \citenamefont {Hwang}, \citenamefont
  {Lambin},\ and\ \citenamefont {Bir\'{o}}}]{Mark2012}%
  \BibitemOpen
  \bibfield  {author} {\bibinfo {author} {\bibfnamefont {G.~I.}\ \bibnamefont
  {M\'{a}rk}}, \bibinfo {author} {\bibfnamefont {P.}~\bibnamefont
  {Vancs\'{o}}}, \bibinfo {author} {\bibfnamefont {C.}~\bibnamefont {Hwang}},
  \bibinfo {author} {\bibfnamefont {P.}~\bibnamefont {Lambin}}, \ and\ \bibinfo
  {author} {\bibfnamefont {L.~P.}\ \bibnamefont {Bir\'{o}}},\ }\href {\doibase
  10.1103/PhysRevB.85.125443} {\bibfield  {journal} {\bibinfo  {journal}
  {Physical Review B}\ }\textbf {\bibinfo {volume} {85}},\ \bibinfo {pages}
  {125443} (\bibinfo {year} {2012})}\BibitemShut {NoStop}%
\bibitem [{\citenamefont {Bergvall}\ and\ \citenamefont
  {L\"{o}fwander}(2013)}]{Bergvall2013}%
  \BibitemOpen
  \bibfield  {author} {\bibinfo {author} {\bibfnamefont {A.}~\bibnamefont
  {Bergvall}}\ and\ \bibinfo {author} {\bibfnamefont {T.}~\bibnamefont
  {L\"{o}fwander}},\ }\href {\doibase 10.1103/PhysRevB.87.205431} {\bibfield
  {journal} {\bibinfo  {journal} {Physical Review B}\ }\textbf {\bibinfo
  {volume} {87}},\ \bibinfo {pages} {205431} (\bibinfo {year}
  {2013})}\BibitemShut {NoStop}%
\bibitem [{\citenamefont {Lawlor}\ \emph {et~al.}(2013)\citenamefont {Lawlor},
  \citenamefont {Power},\ and\ \citenamefont {Ferreira}}]{Lawlor2013}%
  \BibitemOpen
  \bibfield  {author} {\bibinfo {author} {\bibfnamefont {J.~A.}\ \bibnamefont
  {Lawlor}}, \bibinfo {author} {\bibfnamefont {S.~R.}\ \bibnamefont {Power}}, \
  and\ \bibinfo {author} {\bibfnamefont {M.~S.}\ \bibnamefont {Ferreira}},\
  }\href {\doibase 10.1103/PhysRevB.88.205416} {\bibfield  {journal} {\bibinfo
  {journal} {Physical Review B}\ }\textbf {\bibinfo {volume} {88}},\ \bibinfo
  {pages} {205416} (\bibinfo {year} {2013})}\BibitemShut {NoStop}%
\bibitem [{\citenamefont {Lounis}(2014)}]{Lounis2014}%
  \BibitemOpen
  \bibfield  {author} {\bibinfo {author} {\bibfnamefont {S.}~\bibnamefont
  {Lounis}},\ }\href {http://arxiv-web3.library.cornell.edu/abs/1404.0961} {\
  (\bibinfo {year} {2014})},\ \Eprint {http://arxiv.org/abs/1404.0961}
  {arXiv:1404.0961} \BibitemShut {NoStop}%
\bibitem [{\citenamefont {Rutter}\ \emph {et~al.}(2007)\citenamefont {Rutter},
  \citenamefont {Crain}, \citenamefont {Guisinger}, \citenamefont {Li},
  \citenamefont {First},\ and\ \citenamefont {Stroscio}}]{Rutter2007}%
  \BibitemOpen
  \bibfield  {author} {\bibinfo {author} {\bibfnamefont {G.~M.}\ \bibnamefont
  {Rutter}}, \bibinfo {author} {\bibfnamefont {J.~N.}\ \bibnamefont {Crain}},
  \bibinfo {author} {\bibfnamefont {N.~P.}\ \bibnamefont {Guisinger}}, \bibinfo
  {author} {\bibfnamefont {T.}~\bibnamefont {Li}}, \bibinfo {author}
  {\bibfnamefont {P.~N.}\ \bibnamefont {First}}, \ and\ \bibinfo {author}
  {\bibfnamefont {J.~A.}\ \bibnamefont {Stroscio}},\ }\href {\doibase
  10.1126/science.1142882} {\bibfield  {journal} {\bibinfo  {journal}
  {Science}\ }\textbf {\bibinfo {volume} {317}},\ \bibinfo {pages} {219}
  (\bibinfo {year} {2007})}\BibitemShut {NoStop}%
\bibitem [{\citenamefont {Mallet}\ \emph {et~al.}(2007)\citenamefont {Mallet},
  \citenamefont {Varchon}, \citenamefont {Naud}, \citenamefont {Magaud},
  \citenamefont {Berger},\ and\ \citenamefont {Veuillen}}]{Mallet2007}%
  \BibitemOpen
  \bibfield  {author} {\bibinfo {author} {\bibfnamefont {P.}~\bibnamefont
  {Mallet}}, \bibinfo {author} {\bibfnamefont {F.}~\bibnamefont {Varchon}},
  \bibinfo {author} {\bibfnamefont {C.}~\bibnamefont {Naud}}, \bibinfo {author}
  {\bibfnamefont {L.}~\bibnamefont {Magaud}}, \bibinfo {author} {\bibfnamefont
  {C.}~\bibnamefont {Berger}}, \ and\ \bibinfo {author} {\bibfnamefont {J.-Y.}\
  \bibnamefont {Veuillen}},\ }\href {\doibase 10.1103/PhysRevB.76.041403}
  {\bibfield  {journal} {\bibinfo  {journal} {Physical Review B}\ }\textbf
  {\bibinfo {volume} {76}},\ \bibinfo {pages} {041403} (\bibinfo {year}
  {2007})}\BibitemShut {NoStop}%
\bibitem [{\citenamefont {Yang}\ \emph {et~al.}(2010)\citenamefont {Yang},
  \citenamefont {Mayne}, \citenamefont {Boucherit}, \citenamefont {Comtet},
  \citenamefont {Dujardin},\ and\ \citenamefont {Kuk}}]{Yang2010}%
  \BibitemOpen
  \bibfield  {author} {\bibinfo {author} {\bibfnamefont {H.}~\bibnamefont
  {Yang}}, \bibinfo {author} {\bibfnamefont {A.~J.}\ \bibnamefont {Mayne}},
  \bibinfo {author} {\bibfnamefont {M.}~\bibnamefont {Boucherit}}, \bibinfo
  {author} {\bibfnamefont {G.}~\bibnamefont {Comtet}}, \bibinfo {author}
  {\bibfnamefont {G.}~\bibnamefont {Dujardin}}, \ and\ \bibinfo {author}
  {\bibfnamefont {Y.}~\bibnamefont {Kuk}},\ }\href {\doibase 10.1021/nl9038778}
  {\bibfield  {journal} {\bibinfo  {journal} {Nano Letters}\ }\textbf {\bibinfo
  {volume} {10}},\ \bibinfo {pages} {943} (\bibinfo {year} {2010})}\BibitemShut
  {NoStop}%
\bibitem [{\citenamefont {Tapasztó}\ \emph {et~al.}(2012)\citenamefont
  {Tapasztó}, \citenamefont {Nemes-Incze}, \citenamefont {Dobrik},
  \citenamefont {{Jae Yoo}}, \citenamefont {Hwang},\ and\ \citenamefont
  {Biró}}]{Tapaszto2012}%
  \BibitemOpen
  \bibfield  {author} {\bibinfo {author} {\bibfnamefont {L.}~\bibnamefont
  {Tapasztó}}, \bibinfo {author} {\bibfnamefont {P.}~\bibnamefont
  {Nemes-Incze}}, \bibinfo {author} {\bibfnamefont {G.}~\bibnamefont {Dobrik}},
  \bibinfo {author} {\bibfnamefont {K.}~\bibnamefont {{Jae Yoo}}}, \bibinfo
  {author} {\bibfnamefont {C.}~\bibnamefont {Hwang}}, \ and\ \bibinfo {author}
  {\bibfnamefont {L.~P.}\ \bibnamefont {Biró}},\ }\href {\doibase
  10.1063/1.3681375} {\bibfield  {journal} {\bibinfo  {journal} {Applied
  Physics Letters}\ }\textbf {\bibinfo {volume} {100}},\ \bibinfo {pages}
  {053114} (\bibinfo {year} {2012})}\BibitemShut {NoStop}%
\bibitem [{\citenamefont {Xue}\ \emph {et~al.}(2012)\citenamefont {Xue},
  \citenamefont {Sanchez-Yamagishi}, \citenamefont {Watanabe}, \citenamefont
  {Taniguchi}, \citenamefont {Jarillo-Herrero},\ and\ \citenamefont
  {LeRoy}}]{Xue2012}%
  \BibitemOpen
  \bibfield  {author} {\bibinfo {author} {\bibfnamefont {J.}~\bibnamefont
  {Xue}}, \bibinfo {author} {\bibfnamefont {J.}~\bibnamefont
  {Sanchez-Yamagishi}}, \bibinfo {author} {\bibfnamefont {K.}~\bibnamefont
  {Watanabe}}, \bibinfo {author} {\bibfnamefont {T.}~\bibnamefont {Taniguchi}},
  \bibinfo {author} {\bibfnamefont {P.}~\bibnamefont {Jarillo-Herrero}}, \ and\
  \bibinfo {author} {\bibfnamefont {B.~J.}\ \bibnamefont {LeRoy}},\ }\href
  {\doibase 10.1103/PhysRevLett.108.016801} {\bibfield  {journal} {\bibinfo
  {journal} {Physical Review Letters}\ }\textbf {\bibinfo {volume} {108}},\
  \bibinfo {pages} {016801} (\bibinfo {year} {2012})}\BibitemShut {NoStop}%
\bibitem [{\citenamefont {Koepke}\ \emph {et~al.}(2013)\citenamefont {Koepke},
  \citenamefont {Wood}, \citenamefont {Estrada}, \citenamefont {Ong},
  \citenamefont {He}, \citenamefont {Pop},\ and\ \citenamefont
  {Lyding}}]{Koepke2013}%
  \BibitemOpen
  \bibfield  {author} {\bibinfo {author} {\bibfnamefont {J.~C.}\ \bibnamefont
  {Koepke}}, \bibinfo {author} {\bibfnamefont {J.~D.}\ \bibnamefont {Wood}},
  \bibinfo {author} {\bibfnamefont {D.}~\bibnamefont {Estrada}}, \bibinfo
  {author} {\bibfnamefont {Z.-Y.}\ \bibnamefont {Ong}}, \bibinfo {author}
  {\bibfnamefont {K.~T.}\ \bibnamefont {He}}, \bibinfo {author} {\bibfnamefont
  {E.}~\bibnamefont {Pop}}, \ and\ \bibinfo {author} {\bibfnamefont {J.~W.}\
  \bibnamefont {Lyding}},\ }\href {\doibase 10.1021/nn302064p} {\bibfield
  {journal} {\bibinfo  {journal} {ACS Nano}\ }\textbf {\bibinfo {volume} {7}},\
  \bibinfo {pages} {75} (\bibinfo {year} {2013})}\BibitemShut {NoStop}%
\bibitem [{\citenamefont {Hasegawa}\ \emph {et~al.}(2002)\citenamefont
  {Hasegawa}, \citenamefont {Shiraki}, \citenamefont {Tanikawa}, \citenamefont
  {Petersen}, \citenamefont {Hansen}, \citenamefont {B{\o}ggild},\ and\
  \citenamefont {Grey}}]{Hasegawa2002}%
  \BibitemOpen
  \bibfield  {author} {\bibinfo {author} {\bibfnamefont {S.}~\bibnamefont
  {Hasegawa}}, \bibinfo {author} {\bibfnamefont {I.}~\bibnamefont {Shiraki}},
  \bibinfo {author} {\bibfnamefont {T.}~\bibnamefont {Tanikawa}}, \bibinfo
  {author} {\bibfnamefont {C.~L.}\ \bibnamefont {Petersen}}, \bibinfo {author}
  {\bibfnamefont {T.~M.}\ \bibnamefont {Hansen}}, \bibinfo {author}
  {\bibfnamefont {P.}~\bibnamefont {B{\o}ggild}}, \ and\ \bibinfo {author}
  {\bibfnamefont {F.}~\bibnamefont {Grey}},\ }\href
  {http://stacks.iop.org/0953-8984/14/i=35/a=309} {\bibfield  {journal}
  {\bibinfo  {journal} {Journal of Physics: Condensed Matter}\ }\textbf
  {\bibinfo {volume} {14}},\ \bibinfo {pages} {8379} (\bibinfo {year}
  {2002})}\BibitemShut {NoStop}%
\bibitem [{\citenamefont {Kubo}\ \emph {et~al.}(2006)\citenamefont {Kubo},
  \citenamefont {Shingaya}, \citenamefont {Nakaya}, \citenamefont {Aono},\ and\
  \citenamefont {Nakayama}}]{Kubo2006}%
  \BibitemOpen
  \bibfield  {author} {\bibinfo {author} {\bibfnamefont {O.}~\bibnamefont
  {Kubo}}, \bibinfo {author} {\bibfnamefont {Y.}~\bibnamefont {Shingaya}},
  \bibinfo {author} {\bibfnamefont {M.}~\bibnamefont {Nakaya}}, \bibinfo
  {author} {\bibfnamefont {M.}~\bibnamefont {Aono}}, \ and\ \bibinfo {author}
  {\bibfnamefont {T.}~\bibnamefont {Nakayama}},\ }\href {\doibase
  http://dx.doi.org/10.1063/1.2213954} {\bibfield  {journal} {\bibinfo
  {journal} {Applied Physics Letters}\ }\textbf {\bibinfo {volume} {88}},\
  \bibinfo {eid} {254101} (\bibinfo {year} {2006})}\BibitemShut {NoStop}%
\bibitem [{\citenamefont {Jaschinsky}\ \emph {et~al.}(2006)\citenamefont
  {Jaschinsky}, \citenamefont {Coenen}, \citenamefont {Pirug},\ and\
  \citenamefont {Voigtländer}}]{Jaschinsky2006}%
  \BibitemOpen
  \bibfield  {author} {\bibinfo {author} {\bibfnamefont {P.}~\bibnamefont
  {Jaschinsky}}, \bibinfo {author} {\bibfnamefont {P.}~\bibnamefont {Coenen}},
  \bibinfo {author} {\bibfnamefont {G.}~\bibnamefont {Pirug}}, \ and\ \bibinfo
  {author} {\bibfnamefont {B.}~\bibnamefont {Voigtländer}},\ }\href {\doibase
  10.1063/1.2336112} {\bibfield  {journal} {\bibinfo  {journal} {Review of
  Scientific Instruments}\ }\textbf {\bibinfo {volume} {77}},\ \bibinfo {pages}
  {093701} (\bibinfo {year} {2006})}\BibitemShut {NoStop}%
\bibitem [{\citenamefont {Kim}\ \emph {et~al.}(2007)\citenamefont {Kim},
  \citenamefont {Wang}, \citenamefont {Wendelken}, \citenamefont {Weitering},
  \citenamefont {Li},\ and\ \citenamefont {Li}}]{Kim2007}%
  \BibitemOpen
  \bibfield  {author} {\bibinfo {author} {\bibfnamefont {T.-H.}\ \bibnamefont
  {Kim}}, \bibinfo {author} {\bibfnamefont {Z.}~\bibnamefont {Wang}}, \bibinfo
  {author} {\bibfnamefont {J.~F.}\ \bibnamefont {Wendelken}}, \bibinfo {author}
  {\bibfnamefont {H.~H.}\ \bibnamefont {Weitering}}, \bibinfo {author}
  {\bibfnamefont {W.}~\bibnamefont {Li}}, \ and\ \bibinfo {author}
  {\bibfnamefont {A.-P.}\ \bibnamefont {Li}},\ }\href {\doibase
  10.1063/1.2821610} {\bibfield  {journal} {\bibinfo  {journal} {The Review of
  scientific instruments}\ }\textbf {\bibinfo {volume} {78}},\ \bibinfo {pages}
  {123701} (\bibinfo {year} {2007})}\BibitemShut {NoStop}%
\bibitem [{\citenamefont {Baringhaus}\ \emph {et~al.}(2013)\citenamefont
  {Baringhaus}, \citenamefont {Edler}, \citenamefont {Neumann}, \citenamefont
  {Stampfer}, \citenamefont {Forti}, \citenamefont {Starke},\ and\
  \citenamefont {Tegenkamp}}]{Baringhaus2013apl}%
  \BibitemOpen
  \bibfield  {author} {\bibinfo {author} {\bibfnamefont {J.}~\bibnamefont
  {Baringhaus}}, \bibinfo {author} {\bibfnamefont {F.}~\bibnamefont {Edler}},
  \bibinfo {author} {\bibfnamefont {C.}~\bibnamefont {Neumann}}, \bibinfo
  {author} {\bibfnamefont {C.}~\bibnamefont {Stampfer}}, \bibinfo {author}
  {\bibfnamefont {S.}~\bibnamefont {Forti}}, \bibinfo {author} {\bibfnamefont
  {U.}~\bibnamefont {Starke}}, \ and\ \bibinfo {author} {\bibfnamefont
  {C.}~\bibnamefont {Tegenkamp}},\ }\href {\doibase
  http://dx.doi.org/10.1063/1.4821364} {\bibfield  {journal} {\bibinfo
  {journal} {Applied Physics Letters}\ }\textbf {\bibinfo {volume} {103}},\
  \bibinfo {eid} {111604} (\bibinfo {year} {2013})}\BibitemShut {NoStop}%
\bibitem [{\citenamefont {Roychowdhury}\ \emph {et~al.}(2014)\citenamefont
  {Roychowdhury}, \citenamefont {Gubrud}, \citenamefont {Dana}, \citenamefont
  {Lobb}, \citenamefont {Wellstood},\ and\ \citenamefont
  {Dreyer}}]{Roychowdhury2014}%
  \BibitemOpen
  \bibfield  {author} {\bibinfo {author} {\bibfnamefont {A.}~\bibnamefont
  {Roychowdhury}}, \bibinfo {author} {\bibfnamefont {M.~A.}\ \bibnamefont
  {Gubrud}}, \bibinfo {author} {\bibfnamefont {R.}~\bibnamefont {Dana}},
  \bibinfo {author} {\bibfnamefont {C.~J.}\ \bibnamefont {Lobb}}, \bibinfo
  {author} {\bibfnamefont {F.~C.}\ \bibnamefont {Wellstood}}, \ and\ \bibinfo
  {author} {\bibfnamefont {M.}~\bibnamefont {Dreyer}},\ }\href@noop {} {\
  (\bibinfo {year} {2014})},\ \Eprint {http://arxiv.org/abs/arXiv:1311.1855v2}
  {arXiv:1311.1855v2} \BibitemShut {NoStop}%
\bibitem [{\citenamefont {Nakayama}\ \emph {et~al.}(2012)\citenamefont
  {Nakayama}, \citenamefont {Kubo}, \citenamefont {Shingaya}, \citenamefont
  {Higuchi}, \citenamefont {Hasegawa}, \citenamefont {Jiang}, \citenamefont
  {Okuda}, \citenamefont {Kuwahara}, \citenamefont {Takami},\ and\
  \citenamefont {Aono}}]{Nakayama2012}%
  \BibitemOpen
  \bibfield  {author} {\bibinfo {author} {\bibfnamefont {T.}~\bibnamefont
  {Nakayama}}, \bibinfo {author} {\bibfnamefont {O.}~\bibnamefont {Kubo}},
  \bibinfo {author} {\bibfnamefont {Y.}~\bibnamefont {Shingaya}}, \bibinfo
  {author} {\bibfnamefont {S.}~\bibnamefont {Higuchi}}, \bibinfo {author}
  {\bibfnamefont {T.}~\bibnamefont {Hasegawa}}, \bibinfo {author}
  {\bibfnamefont {C.-S.}\ \bibnamefont {Jiang}}, \bibinfo {author}
  {\bibfnamefont {T.}~\bibnamefont {Okuda}}, \bibinfo {author} {\bibfnamefont
  {Y.}~\bibnamefont {Kuwahara}}, \bibinfo {author} {\bibfnamefont
  {K.}~\bibnamefont {Takami}}, \ and\ \bibinfo {author} {\bibfnamefont
  {M.}~\bibnamefont {Aono}},\ }\href {\doibase 10.1002/adma.201200257}
  {\bibfield  {journal} {\bibinfo  {journal} {Advanced materials}\ }\textbf
  {\bibinfo {volume} {24}},\ \bibinfo {pages} {1675} (\bibinfo {year}
  {2012})}\BibitemShut {NoStop}%
\bibitem [{\citenamefont {Li}\ \emph {et~al.}(2013)\citenamefont {Li},
  \citenamefont {Clark}, \citenamefont {Zhang},\ and\ \citenamefont
  {Baddorf}}]{Li2013}%
  \BibitemOpen
  \bibfield  {author} {\bibinfo {author} {\bibfnamefont {A.-P.}\ \bibnamefont
  {Li}}, \bibinfo {author} {\bibfnamefont {K.~W.}\ \bibnamefont {Clark}},
  \bibinfo {author} {\bibfnamefont {X.-G.}\ \bibnamefont {Zhang}}, \ and\
  \bibinfo {author} {\bibfnamefont {A.~P.}\ \bibnamefont {Baddorf}},\ }\href
  {\doibase 10.1002/adfm.201203423} {\bibfield  {journal} {\bibinfo  {journal}
  {Advanced Functional Materials}\ }\textbf {\bibinfo {volume} {23}},\ \bibinfo
  {pages} {2509} (\bibinfo {year} {2013})}\BibitemShut {NoStop}%
\bibitem [{\citenamefont {Baringhaus}\ \emph {et~al.}(2014)\citenamefont
  {Baringhaus}, \citenamefont {Ruan}, \citenamefont {Edler}, \citenamefont
  {Tejeda}, \citenamefont {Sicot}, \citenamefont {Li}, \citenamefont {Jiang},
  \citenamefont {Conrad}, \citenamefont {Berger}, \citenamefont {Tegenkamp},\
  and\ \citenamefont {de~Heer}}]{Baringhaus2014}%
  \BibitemOpen
  \bibfield  {author} {\bibinfo {author} {\bibfnamefont {J.}~\bibnamefont
  {Baringhaus}}, \bibinfo {author} {\bibfnamefont {M.}~\bibnamefont {Ruan}},
  \bibinfo {author} {\bibfnamefont {F.}~\bibnamefont {Edler}}, \bibinfo
  {author} {\bibfnamefont {A.}~\bibnamefont {Tejeda}}, \bibinfo {author}
  {\bibfnamefont {M.}~\bibnamefont {Sicot}}, \bibinfo {author} {\bibfnamefont
  {A.-P.}\ \bibnamefont {Li}}, \bibinfo {author} {\bibfnamefont
  {Z.}~\bibnamefont {Jiang}}, \bibinfo {author} {\bibfnamefont {E.~H.}\
  \bibnamefont {Conrad}}, \bibinfo {author} {\bibfnamefont {C.}~\bibnamefont
  {Berger}}, \bibinfo {author} {\bibfnamefont {C.}~\bibnamefont {Tegenkamp}}, \
  and\ \bibinfo {author} {\bibfnamefont {W.~A.}\ \bibnamefont {de~Heer}},\
  }\href {\doibase 10.1038/nature12952} {\bibfield  {journal} {\bibinfo
  {journal} {Nature}\ }\textbf {\bibinfo {volume} {506}},\ \bibinfo {pages}
  {349} (\bibinfo {year} {2014})}\BibitemShut {NoStop}%
\bibitem [{\citenamefont {Kanagawa}\ \emph {et~al.}(2003)\citenamefont
  {Kanagawa}, \citenamefont {Hobara}, \citenamefont {Matsuda}, \citenamefont
  {Tanikawa}, \citenamefont {Natori},\ and\ \citenamefont
  {Hasegawa}}]{Kanagawa2003}%
  \BibitemOpen
  \bibfield  {author} {\bibinfo {author} {\bibfnamefont {T.}~\bibnamefont
  {Kanagawa}}, \bibinfo {author} {\bibfnamefont {R.}~\bibnamefont {Hobara}},
  \bibinfo {author} {\bibfnamefont {I.}~\bibnamefont {Matsuda}}, \bibinfo
  {author} {\bibfnamefont {T.}~\bibnamefont {Tanikawa}}, \bibinfo {author}
  {\bibfnamefont {A.}~\bibnamefont {Natori}}, \ and\ \bibinfo {author}
  {\bibfnamefont {S.}~\bibnamefont {Hasegawa}},\ }\href {\doibase
  10.1103/PhysRevLett.91.036805} {\bibfield  {journal} {\bibinfo  {journal}
  {Phys. Rev. Lett.}\ }\textbf {\bibinfo {volume} {91}},\ \bibinfo {pages}
  {036805} (\bibinfo {year} {2003})}\BibitemShut {NoStop}%
\bibitem [{\citenamefont {Cherepanov}\ \emph {et~al.}(2012)\citenamefont
  {Cherepanov}, \citenamefont {Zubkov}, \citenamefont {Junker}, \citenamefont
  {Korte}, \citenamefont {Blab}, \citenamefont {Coenen},\ and\ \citenamefont
  {Voigtl\"{a}nder}}]{Cherepanov2012}%
  \BibitemOpen
  \bibfield  {author} {\bibinfo {author} {\bibfnamefont {V.}~\bibnamefont
  {Cherepanov}}, \bibinfo {author} {\bibfnamefont {E.}~\bibnamefont {Zubkov}},
  \bibinfo {author} {\bibfnamefont {H.}~\bibnamefont {Junker}}, \bibinfo
  {author} {\bibfnamefont {S.}~\bibnamefont {Korte}}, \bibinfo {author}
  {\bibfnamefont {M.}~\bibnamefont {Blab}}, \bibinfo {author} {\bibfnamefont
  {P.}~\bibnamefont {Coenen}}, \ and\ \bibinfo {author} {\bibfnamefont
  {B.}~\bibnamefont {Voigtl\"{a}nder}},\ }\href {\doibase 10.1063/1.3694990}
  {\bibfield  {journal} {\bibinfo  {journal} {Review of scientific
  instruments}\ }\textbf {\bibinfo {volume} {83}},\ \bibinfo {pages} {033707}
  (\bibinfo {year} {2012})}\BibitemShut {NoStop}%
\bibitem [{\citenamefont {Watanabe}\ \emph {et~al.}(2001)\citenamefont
  {Watanabe}, \citenamefont {Manabe}, \citenamefont {Shigematsu},\ and\
  \citenamefont {Shimizu}}]{Watanabe2001}%
  \BibitemOpen
  \bibfield  {author} {\bibinfo {author} {\bibfnamefont {H.}~\bibnamefont
  {Watanabe}}, \bibinfo {author} {\bibfnamefont {C.}~\bibnamefont {Manabe}},
  \bibinfo {author} {\bibfnamefont {T.}~\bibnamefont {Shigematsu}}, \ and\
  \bibinfo {author} {\bibfnamefont {M.}~\bibnamefont {Shimizu}},\ }\href@noop
  {} {\bibfield  {journal} {\bibinfo  {journal} {Applied Physics Letters}\
  }\textbf {\bibinfo {volume} {78}} (\bibinfo {year} {2001})}\BibitemShut
  {NoStop}%
\bibitem [{\citenamefont {Clark}\ \emph {et~al.}(2013)\citenamefont {Clark},
  \citenamefont {Zhang}, \citenamefont {Vlassiouk}, \citenamefont {He},
  \citenamefont {Feenstra},\ and\ \citenamefont {Li}}]{Clark2013}%
  \BibitemOpen
  \bibfield  {author} {\bibinfo {author} {\bibfnamefont {K.~W.}\ \bibnamefont
  {Clark}}, \bibinfo {author} {\bibfnamefont {X.-G.}\ \bibnamefont {Zhang}},
  \bibinfo {author} {\bibfnamefont {I.~V.}\ \bibnamefont {Vlassiouk}}, \bibinfo
  {author} {\bibfnamefont {G.}~\bibnamefont {He}}, \bibinfo {author}
  {\bibfnamefont {R.~M.}\ \bibnamefont {Feenstra}}, \ and\ \bibinfo {author}
  {\bibfnamefont {A.-P.}\ \bibnamefont {Li}},\ }\href {\doibase
  10.1021/nn403056k} {\bibfield  {journal} {\bibinfo  {journal} {ACS nano}\
  }\textbf {\bibinfo {volume} {7}},\ \bibinfo {pages} {7956} (\bibinfo {year}
  {2013})}\BibitemShut {NoStop}%
\bibitem [{\citenamefont {Clark}\ \emph {et~al.}(2014)\citenamefont {Clark},
  \citenamefont {Zhang}, \citenamefont {Gu}, \citenamefont {Park},
  \citenamefont {He}, \citenamefont {Feenstra},\ and\ \citenamefont
  {Li}}]{Clark2014}%
  \BibitemOpen
  \bibfield  {author} {\bibinfo {author} {\bibfnamefont {K.~W.}\ \bibnamefont
  {Clark}}, \bibinfo {author} {\bibfnamefont {X.-G.}\ \bibnamefont {Zhang}},
  \bibinfo {author} {\bibfnamefont {G.}~\bibnamefont {Gu}}, \bibinfo {author}
  {\bibfnamefont {J.}~\bibnamefont {Park}}, \bibinfo {author} {\bibfnamefont
  {G.}~\bibnamefont {He}}, \bibinfo {author} {\bibfnamefont {R.~M.}\
  \bibnamefont {Feenstra}}, \ and\ \bibinfo {author} {\bibfnamefont {A.-P.}\
  \bibnamefont {Li}},\ }\href {\doibase 10.1103/PhysRevX.4.011021} {\bibfield
  {journal} {\bibinfo  {journal} {Physical Review X}\ }\textbf {\bibinfo
  {volume} {4}},\ \bibinfo {pages} {011021} (\bibinfo {year}
  {2014})}\BibitemShut {NoStop}%
\bibitem [{\citenamefont {Kim}\ \emph {et~al.}(2010)\citenamefont {Kim},
  \citenamefont {Zhang}, \citenamefont {Nicholson}, \citenamefont {Evans},
  \citenamefont {Kulkarni}, \citenamefont {Radhakrishnan}, \citenamefont
  {Kenik},\ and\ \citenamefont {Li}}]{Kim2010}%
  \BibitemOpen
  \bibfield  {author} {\bibinfo {author} {\bibfnamefont {T.-H.}\ \bibnamefont
  {Kim}}, \bibinfo {author} {\bibfnamefont {X.-G.}\ \bibnamefont {Zhang}},
  \bibinfo {author} {\bibfnamefont {D.~M.}\ \bibnamefont {Nicholson}}, \bibinfo
  {author} {\bibfnamefont {B.~M.}\ \bibnamefont {Evans}}, \bibinfo {author}
  {\bibfnamefont {N.~S.}\ \bibnamefont {Kulkarni}}, \bibinfo {author}
  {\bibfnamefont {B.}~\bibnamefont {Radhakrishnan}}, \bibinfo {author}
  {\bibfnamefont {E.~A.}\ \bibnamefont {Kenik}}, \ and\ \bibinfo {author}
  {\bibfnamefont {A.-P.}\ \bibnamefont {Li}},\ }\href {\doibase
  10.1021/nl101734h} {\bibfield  {journal} {\bibinfo  {journal} {Nano Letters}\
  }\textbf {\bibinfo {volume} {10}},\ \bibinfo {pages} {3096} (\bibinfo {year}
  {2010})}\BibitemShut {NoStop}%
\bibitem [{\citenamefont {Eder}\ \emph {et~al.}(2013)\citenamefont {Eder},
  \citenamefont {Kotakoski}, \citenamefont {Holzweber}, \citenamefont
  {Mangler}, \citenamefont {Skakalova},\ and\ \citenamefont
  {Meyer}}]{Eder2013}%
  \BibitemOpen
  \bibfield  {author} {\bibinfo {author} {\bibfnamefont {F.~R.}\ \bibnamefont
  {Eder}}, \bibinfo {author} {\bibfnamefont {J.}~\bibnamefont {Kotakoski}},
  \bibinfo {author} {\bibfnamefont {K.}~\bibnamefont {Holzweber}}, \bibinfo
  {author} {\bibfnamefont {C.}~\bibnamefont {Mangler}}, \bibinfo {author}
  {\bibfnamefont {V.}~\bibnamefont {Skakalova}}, \ and\ \bibinfo {author}
  {\bibfnamefont {J.~C.}\ \bibnamefont {Meyer}},\ }\href {\doibase
  10.1021/nl3042799} {\bibfield  {journal} {\bibinfo  {journal} {Nano Letters}\
  }\textbf {\bibinfo {volume} {13}},\ \bibinfo {pages} {1934} (\bibinfo {year}
  {2013})}\BibitemShut {NoStop}%
\bibitem [{\citenamefont {Ji}\ \emph {et~al.}(2012)\citenamefont {Ji},
  \citenamefont {Hannon}, \citenamefont {Tromp}, \citenamefont {Perebeinos},
  \citenamefont {Tersoff},\ and\ \citenamefont {Ross}}]{Ji2012}%
  \BibitemOpen
  \bibfield  {author} {\bibinfo {author} {\bibfnamefont {S.-H.}\ \bibnamefont
  {Ji}}, \bibinfo {author} {\bibfnamefont {J.~B.}\ \bibnamefont {Hannon}},
  \bibinfo {author} {\bibfnamefont {R.~M.}\ \bibnamefont {Tromp}}, \bibinfo
  {author} {\bibfnamefont {V.}~\bibnamefont {Perebeinos}}, \bibinfo {author}
  {\bibfnamefont {J.}~\bibnamefont {Tersoff}}, \ and\ \bibinfo {author}
  {\bibfnamefont {F.~M.}\ \bibnamefont {Ross}},\ }\href {\doibase
  10.1038/nmat3170} {\bibfield  {journal} {\bibinfo  {journal} {Nature
  materials}\ }\textbf {\bibinfo {volume} {11}},\ \bibinfo {pages} {114}
  (\bibinfo {year} {2012})}\BibitemShut {NoStop}%
\bibitem [{\citenamefont {Sutter}\ \emph {et~al.}(2008)\citenamefont {Sutter},
  \citenamefont {Flege},\ and\ \citenamefont {Sutter}}]{Sutter2008}%
  \BibitemOpen
  \bibfield  {author} {\bibinfo {author} {\bibfnamefont {P.~W.}\ \bibnamefont
  {Sutter}}, \bibinfo {author} {\bibfnamefont {J.-I.}\ \bibnamefont {Flege}}, \
  and\ \bibinfo {author} {\bibfnamefont {E.~A.}\ \bibnamefont {Sutter}},\
  }\href {\doibase 10.1038/nmat2166} {\bibfield  {journal} {\bibinfo  {journal}
  {Nature Materials}\ }\textbf {\bibinfo {volume} {7}},\ \bibinfo {pages} {406}
  (\bibinfo {year} {2008})}\BibitemShut {NoStop}%
\bibitem [{\citenamefont {Settnes}\ \emph {et~al.}(2014)\citenamefont
  {Settnes}, \citenamefont {Power}, \citenamefont {Petersen},\ and\
  \citenamefont {Jauho}}]{Settnes2014}%
  \BibitemOpen
  \bibfield  {author} {\bibinfo {author} {\bibfnamefont {M.}~\bibnamefont
  {Settnes}}, \bibinfo {author} {\bibfnamefont {S.~R.}\ \bibnamefont {Power}},
  \bibinfo {author} {\bibfnamefont {D.~H.}\ \bibnamefont {Petersen}}, \ and\
  \bibinfo {author} {\bibfnamefont {A.-P.}\ \bibnamefont {Jauho}},\ }\href
  {\doibase 10.1103/PhysRevLett.112.096801} {\bibfield  {journal} {\bibinfo
  {journal} {Phys. Rev. Lett.}\ }\textbf {\bibinfo {volume} {112}},\ \bibinfo
  {pages} {096801} (\bibinfo {year} {2014})}\BibitemShut {NoStop}%
\bibitem [{\citenamefont {Datta}(1997)}]{DattaBook}%
  \BibitemOpen
  \bibfield  {author} {\bibinfo {author} {\bibfnamefont {S.}~\bibnamefont
  {Datta}},\ }\href@noop {} {\emph {\bibinfo {title} {Electronic Transport in
  Mesoscopic Systems}}}\ (\bibinfo  {publisher} {Cambridge University Press},\
  \bibinfo {year} {1997})\BibitemShut {NoStop}%
\bibitem [{\citenamefont {Reich}\ \emph {et~al.}(2002)\citenamefont {Reich},
  \citenamefont {Maultzsch}, \citenamefont {Thomsen},\ and\ \citenamefont
  {Ordej\'{o}n}}]{Reich2002}%
  \BibitemOpen
  \bibfield  {author} {\bibinfo {author} {\bibfnamefont {S.}~\bibnamefont
  {Reich}}, \bibinfo {author} {\bibfnamefont {J.}~\bibnamefont {Maultzsch}},
  \bibinfo {author} {\bibfnamefont {C.}~\bibnamefont {Thomsen}}, \ and\
  \bibinfo {author} {\bibfnamefont {P.}~\bibnamefont {Ordej\'{o}n}},\ }\href
  {\doibase 10.1103/PhysRevB.66.035412} {\bibfield  {journal} {\bibinfo
  {journal} {Physical Review B}\ }\textbf {\bibinfo {volume} {66}},\ \bibinfo
  {pages} {035412} (\bibinfo {year} {2002})}\BibitemShut {NoStop}%
\bibitem [{\citenamefont {Power}\ and\ \citenamefont
  {Ferreira}(2011)}]{Power2011}%
  \BibitemOpen
  \bibfield  {author} {\bibinfo {author} {\bibfnamefont {S.~R.}\ \bibnamefont
  {Power}}\ and\ \bibinfo {author} {\bibfnamefont {M.~S.}\ \bibnamefont
  {Ferreira}},\ }\href {\doibase 10.1103/PhysRevB.83.155432} {\bibfield
  {journal} {\bibinfo  {journal} {Physical Review B}\ }\textbf {\bibinfo
  {volume} {83}},\ \bibinfo {pages} {155432} (\bibinfo {year}
  {2011})}\BibitemShut {NoStop}%
\bibitem [{\citenamefont {Lherbier}\ \emph {et~al.}(2012)\citenamefont
  {Lherbier}, \citenamefont {Dubois}, \citenamefont {Declerck}, \citenamefont
  {Niquet}, \citenamefont {Roche},\ and\ \citenamefont
  {Charlier}}]{Lherbier2012}%
  \BibitemOpen
  \bibfield  {author} {\bibinfo {author} {\bibfnamefont {A.}~\bibnamefont
  {Lherbier}}, \bibinfo {author} {\bibfnamefont {S.~M.~M.}\ \bibnamefont
  {Dubois}}, \bibinfo {author} {\bibfnamefont {X.}~\bibnamefont {Declerck}},
  \bibinfo {author} {\bibfnamefont {Y.-M.}\ \bibnamefont {Niquet}}, \bibinfo
  {author} {\bibfnamefont {S.}~\bibnamefont {Roche}}, \ and\ \bibinfo {author}
  {\bibfnamefont {J.-C.}\ \bibnamefont {Charlier}},\ }\href {\doibase
  10.1103/PhysRevB.86.075402} {\bibfield  {journal} {\bibinfo  {journal}
  {Physical Review B}\ }\textbf {\bibinfo {volume} {86}},\ \bibinfo {pages}
  {075402} (\bibinfo {year} {2012})}\BibitemShut {NoStop}%
\bibitem [{\citenamefont {Torres}\ \emph {et~al.}(2014)\citenamefont {Torres},
  \citenamefont {Roche},\ and\ \citenamefont {Charlier}}]{RocheBook}%
  \BibitemOpen
  \bibfield  {author} {\bibinfo {author} {\bibfnamefont {L.~E. F.~F.}\
  \bibnamefont {Torres}}, \bibinfo {author} {\bibfnamefont {S.}~\bibnamefont
  {Roche}}, \ and\ \bibinfo {author} {\bibfnamefont {J.-C.}\ \bibnamefont
  {Charlier}},\ }\href@noop {} {\emph {\bibinfo {title} {Introduction to
  Graphene-Based Nanomaterials}}}\ (\bibinfo  {publisher} {Cambridge University
  Press},\ \bibinfo {year} {2014})\BibitemShut {NoStop}%
\bibitem [{\citenamefont {Haug}\ and\ \citenamefont {Jauho}(2008)}]{AnttiBook}%
  \BibitemOpen
  \bibfield  {author} {\bibinfo {author} {\bibfnamefont {H.}~\bibnamefont
  {Haug}}\ and\ \bibinfo {author} {\bibfnamefont {A.-P.}\ \bibnamefont
  {Jauho}},\ }\href@noop {} {\emph {\bibinfo {title} {Quantum kinetics in
  transport and optics of semiconductors}}}\ (\bibinfo  {publisher}
  {Springer},\ \bibinfo {year} {2008})\BibitemShut {NoStop}%
\bibitem [{\citenamefont {Economou}(2005)}]{ecoBook}%
  \BibitemOpen
  \bibfield  {author} {\bibinfo {author} {\bibfnamefont {E.~N.}\ \bibnamefont
  {Economou}},\ }\href@noop {} {\emph {\bibinfo {title} {Green's functions in
  quantum physics}}}\ (\bibinfo  {publisher} {Springer},\ \bibinfo {year}
  {2005})\BibitemShut {NoStop}%
\bibitem [{\citenamefont {Tersoff}\ and\ \citenamefont
  {Hamann}(1983)}]{Tersoff1983}%
  \BibitemOpen
  \bibfield  {author} {\bibinfo {author} {\bibfnamefont {J.}~\bibnamefont
  {Tersoff}}\ and\ \bibinfo {author} {\bibfnamefont {D.~R.}\ \bibnamefont
  {Hamann}},\ }\href {\doibase 10.1103/PhysRevLett.50.1998} {\bibfield
  {journal} {\bibinfo  {journal} {Physical Review Letters}\ }\textbf {\bibinfo
  {volume} {50}},\ \bibinfo {pages} {1998} (\bibinfo {year}
  {1983})}\BibitemShut {NoStop}%
\bibitem [{\citenamefont {Meunier}\ and\ \citenamefont
  {Lambin}(1998)}]{Meunier1998}%
  \BibitemOpen
  \bibfield  {author} {\bibinfo {author} {\bibfnamefont {V.}~\bibnamefont
  {Meunier}}\ and\ \bibinfo {author} {\bibfnamefont {P.}~\bibnamefont
  {Lambin}},\ }\href {\doibase 10.1103/PhysRevLett.81.5588} {\bibfield
  {journal} {\bibinfo  {journal} {Physical Review Letters}\ }\textbf {\bibinfo
  {volume} {81}},\ \bibinfo {pages} {5588} (\bibinfo {year}
  {1998})}\BibitemShut {NoStop}%
\bibitem [{\citenamefont {Amara}\ \emph {et~al.}(2007)\citenamefont {Amara},
  \citenamefont {Latil}, \citenamefont {Meunier}, \citenamefont {Lambin},\ and\
  \citenamefont {Charlier}}]{Amara2007}%
  \BibitemOpen
  \bibfield  {author} {\bibinfo {author} {\bibfnamefont {H.}~\bibnamefont
  {Amara}}, \bibinfo {author} {\bibfnamefont {S.}~\bibnamefont {Latil}},
  \bibinfo {author} {\bibfnamefont {V.}~\bibnamefont {Meunier}}, \bibinfo
  {author} {\bibfnamefont {P.}~\bibnamefont {Lambin}}, \ and\ \bibinfo {author}
  {\bibfnamefont {J.-C.}\ \bibnamefont {Charlier}},\ }\href {\doibase
  10.1103/PhysRevB.76.115423} {\bibfield  {journal} {\bibinfo  {journal}
  {Physical Review B}\ }\textbf {\bibinfo {volume} {76}},\ \bibinfo {pages}
  {115423} (\bibinfo {year} {2007})}\BibitemShut {NoStop}%
\bibitem [{\citenamefont {Zhang}\ \emph {et~al.}(2008)\citenamefont {Zhang},
  \citenamefont {Brar}, \citenamefont {Wang}, \citenamefont {Girit},
  \citenamefont {Yayon}, \citenamefont {Panlasigui}, \citenamefont {Zettl},\
  and\ \citenamefont {Crommie}}]{Zhang2008}%
  \BibitemOpen
  \bibfield  {author} {\bibinfo {author} {\bibfnamefont {Y.}~\bibnamefont
  {Zhang}}, \bibinfo {author} {\bibfnamefont {V.~W.}\ \bibnamefont {Brar}},
  \bibinfo {author} {\bibfnamefont {F.}~\bibnamefont {Wang}}, \bibinfo {author}
  {\bibfnamefont {C.}~\bibnamefont {Girit}}, \bibinfo {author} {\bibfnamefont
  {Y.}~\bibnamefont {Yayon}}, \bibinfo {author} {\bibfnamefont
  {M.}~\bibnamefont {Panlasigui}}, \bibinfo {author} {\bibfnamefont
  {A.}~\bibnamefont {Zettl}}, \ and\ \bibinfo {author} {\bibfnamefont {M.~F.}\
  \bibnamefont {Crommie}},\ }\href {\doibase 10.1038/nphys1022} {\bibfield
  {journal} {\bibinfo  {journal} {Nature Physics}\ }\textbf {\bibinfo {volume}
  {4}},\ \bibinfo {pages} {627} (\bibinfo {year} {2008})}\BibitemShut {NoStop}%
\bibitem [{\citenamefont {Fasolino}\ \emph {et~al.}(2007)\citenamefont
  {Fasolino}, \citenamefont {Los},\ and\ \citenamefont
  {Katsnelson}}]{Fasolino2007}%
  \BibitemOpen
  \bibfield  {author} {\bibinfo {author} {\bibfnamefont {A.}~\bibnamefont
  {Fasolino}}, \bibinfo {author} {\bibfnamefont {J.~H.}\ \bibnamefont {Los}}, \
  and\ \bibinfo {author} {\bibfnamefont {M.~I.}\ \bibnamefont {Katsnelson}},\
  }\href {\doibase 10.1038/nmat2011} {\bibfield  {journal} {\bibinfo  {journal}
  {Nature Materials}\ }\textbf {\bibinfo {volume} {6}},\ \bibinfo {pages} {858}
  (\bibinfo {year} {2007})}\BibitemShut {NoStop}%
\bibitem [{\citenamefont {Robinson}\ \emph {et~al.}(2008)\citenamefont
  {Robinson}, \citenamefont {Schomerus}, \citenamefont {Oroszl\'any},\ and\
  \citenamefont {Fal'ko}}]{Robinson2008}%
  \BibitemOpen
  \bibfield  {author} {\bibinfo {author} {\bibfnamefont {J.~P.}\ \bibnamefont
  {Robinson}}, \bibinfo {author} {\bibfnamefont {H.}~\bibnamefont {Schomerus}},
  \bibinfo {author} {\bibfnamefont {L.}~\bibnamefont {Oroszl\'any}}, \ and\
  \bibinfo {author} {\bibfnamefont {V.~I.}\ \bibnamefont {Fal'ko}},\ }\href
  {\doibase 10.1103/PhysRevLett.101.196803} {\bibfield  {journal} {\bibinfo
  {journal} {Phys. Rev. Lett.}\ }\textbf {\bibinfo {volume} {101}},\ \bibinfo
  {pages} {196803} (\bibinfo {year} {2008})}\BibitemShut {NoStop}%
\bibitem [{\citenamefont {Wehling}\ \emph {et~al.}(2010)\citenamefont
  {Wehling}, \citenamefont {Yuan}, \citenamefont {Lichtenstein}, \citenamefont
  {Geim},\ and\ \citenamefont {Katsnelson}}]{Wehling2010}%
  \BibitemOpen
  \bibfield  {author} {\bibinfo {author} {\bibfnamefont {T.~O.}\ \bibnamefont
  {Wehling}}, \bibinfo {author} {\bibfnamefont {S.}~\bibnamefont {Yuan}},
  \bibinfo {author} {\bibfnamefont {A.~I.}\ \bibnamefont {Lichtenstein}},
  \bibinfo {author} {\bibfnamefont {A.~K.}\ \bibnamefont {Geim}}, \ and\
  \bibinfo {author} {\bibfnamefont {M.~I.}\ \bibnamefont {Katsnelson}},\ }\href
  {\doibase 10.1103/PhysRevLett.105.056802} {\bibfield  {journal} {\bibinfo
  {journal} {Phys. Rev. Lett.}\ }\textbf {\bibinfo {volume} {105}},\ \bibinfo
  {pages} {056802} (\bibinfo {year} {2010})}\BibitemShut {NoStop}%
\bibitem [{\citenamefont {Uchoa}\ \emph {et~al.}(2009)\citenamefont {Uchoa},
  \citenamefont {Yang}, \citenamefont {Tsai}, \citenamefont {Peres},\ and\
  \citenamefont {{Castro Neto}}}]{Uchoa2009}%
  \BibitemOpen
  \bibfield  {author} {\bibinfo {author} {\bibfnamefont {B.}~\bibnamefont
  {Uchoa}}, \bibinfo {author} {\bibfnamefont {L.}~\bibnamefont {Yang}},
  \bibinfo {author} {\bibfnamefont {S.~W.}\ \bibnamefont {Tsai}}, \bibinfo
  {author} {\bibfnamefont {N.~M.~R.}\ \bibnamefont {Peres}}, \ and\ \bibinfo
  {author} {\bibfnamefont {A.~H.}\ \bibnamefont {{Castro Neto}}},\ }\href
  {\doibase 10.1103/PhysRevLett.103.206804} {\bibfield  {journal} {\bibinfo
  {journal} {Physical Review Letters}\ }\textbf {\bibinfo {volume} {103}},\
  \bibinfo {pages} {206804} (\bibinfo {year} {2009})}\BibitemShut {NoStop}%
\bibitem [{\citenamefont {Fano}(1961)}]{Fano1961}%
  \BibitemOpen
  \bibfield  {author} {\bibinfo {author} {\bibfnamefont {U.}~\bibnamefont
  {Fano}},\ }\href {\doibase 10.1103/PhysRev.124.1866} {\bibfield  {journal}
  {\bibinfo  {journal} {Phys. Rev.}\ }\textbf {\bibinfo {volume} {124}},\
  \bibinfo {pages} {1866} (\bibinfo {year} {1961})}\BibitemShut {NoStop}%
\bibitem [{\citenamefont {Schedin}\ \emph {et~al.}(2007)\citenamefont
  {Schedin}, \citenamefont {Geim}, \citenamefont {Morozov}, \citenamefont
  {Hill}, \citenamefont {Blake}, \citenamefont {Katsnelson},\ and\
  \citenamefont {Novoselov}}]{Schedin2007}%
  \BibitemOpen
  \bibfield  {author} {\bibinfo {author} {\bibfnamefont {F.}~\bibnamefont
  {Schedin}}, \bibinfo {author} {\bibfnamefont {A.~K.}\ \bibnamefont {Geim}},
  \bibinfo {author} {\bibfnamefont {S.~V.}\ \bibnamefont {Morozov}}, \bibinfo
  {author} {\bibfnamefont {E.~W.}\ \bibnamefont {Hill}}, \bibinfo {author}
  {\bibfnamefont {P.}~\bibnamefont {Blake}}, \bibinfo {author} {\bibfnamefont
  {M.~I.}\ \bibnamefont {Katsnelson}}, \ and\ \bibinfo {author} {\bibfnamefont
  {K.~S.}\ \bibnamefont {Novoselov}},\ }\href {\doibase 10.1038/nmat1967}
  {\bibfield  {journal} {\bibinfo  {journal} {Nature Materials}\ }\textbf
  {\bibinfo {volume} {6}},\ \bibinfo {pages} {652} (\bibinfo {year}
  {2007})}\BibitemShut {NoStop}%
\bibitem [{\citenamefont {Cagliani}\ \emph {et~al.}(2014)\citenamefont
  {Cagliani}, \citenamefont {Mackenzie}, \citenamefont {Tschammer},
  \citenamefont {Pizzocchero}, \citenamefont {Almdal},\ and\ \citenamefont
  {B{\o}ggild}}]{Cagliani2014}%
  \BibitemOpen
  \bibfield  {author} {\bibinfo {author} {\bibfnamefont {A.}~\bibnamefont
  {Cagliani}}, \bibinfo {author} {\bibfnamefont {D.}~\bibnamefont {Mackenzie}},
  \bibinfo {author} {\bibfnamefont {L.~K.}\ \bibnamefont {Tschammer}}, \bibinfo
  {author} {\bibfnamefont {F.}~\bibnamefont {Pizzocchero}}, \bibinfo {author}
  {\bibfnamefont {K.}~\bibnamefont {Almdal}}, \ and\ \bibinfo {author}
  {\bibfnamefont {P.}~\bibnamefont {B{\o}ggild}},\ }\href
  {http://arxiv.org/abs/1403.4791} {\  (\bibinfo {year} {2014})},\ \Eprint
  {http://arxiv.org/abs/1403.4791} {arXiv:1403.4791} \BibitemShut {NoStop}%
\bibitem [{\citenamefont {F\"{u}rst}\ \emph {et~al.}(2009)\citenamefont
  {F\"{u}rst}, \citenamefont {Pedersen}, \citenamefont {Flindt}, \citenamefont
  {Mortensen}, \citenamefont {Brandbyge}, \citenamefont {Pedersen},\ and\
  \citenamefont {Jauho}}]{Furst2009}%
  \BibitemOpen
  \bibfield  {author} {\bibinfo {author} {\bibfnamefont {J.~A.}\ \bibnamefont
  {F\"{u}rst}}, \bibinfo {author} {\bibfnamefont {J.~G.}\ \bibnamefont
  {Pedersen}}, \bibinfo {author} {\bibfnamefont {C.}~\bibnamefont {Flindt}},
  \bibinfo {author} {\bibfnamefont {N.~A.}\ \bibnamefont {Mortensen}}, \bibinfo
  {author} {\bibfnamefont {M.}~\bibnamefont {Brandbyge}}, \bibinfo {author}
  {\bibfnamefont {T.~G.}\ \bibnamefont {Pedersen}}, \ and\ \bibinfo {author}
  {\bibfnamefont {A.~P.}\ \bibnamefont {Jauho}},\ }\href {\doibase
  10.1088/1367-2630/11/9/095020} {\bibfield  {journal} {\bibinfo  {journal}
  {New Journal of Physics}\ }\textbf {\bibinfo {volume} {11}},\ \bibinfo
  {pages} {095020} (\bibinfo {year} {2009})}\BibitemShut {NoStop}%
\bibitem [{\citenamefont {Pedersen}\ \emph {et~al.}(2008)\citenamefont
  {Pedersen}, \citenamefont {Flindt}, \citenamefont {Pedersen}, \citenamefont
  {Mortensen}, \citenamefont {Jauho},\ and\ \citenamefont
  {Pedersen}}]{Pedersen2008}%
  \BibitemOpen
  \bibfield  {author} {\bibinfo {author} {\bibfnamefont {T.~G.}\ \bibnamefont
  {Pedersen}}, \bibinfo {author} {\bibfnamefont {C.}~\bibnamefont {Flindt}},
  \bibinfo {author} {\bibfnamefont {J.~G.}\ \bibnamefont {Pedersen}}, \bibinfo
  {author} {\bibfnamefont {N.~A.}\ \bibnamefont {Mortensen}}, \bibinfo {author}
  {\bibfnamefont {A.-P.}\ \bibnamefont {Jauho}}, \ and\ \bibinfo {author}
  {\bibfnamefont {K.}~\bibnamefont {Pedersen}},\ }\href {\doibase
  10.1103/PhysRevLett.100.136804} {\bibfield  {journal} {\bibinfo  {journal}
  {Physical Review Letters}\ }\textbf {\bibinfo {volume} {100}},\ \bibinfo
  {pages} {136804} (\bibinfo {year} {2008})}\BibitemShut {NoStop}%
\bibitem [{\citenamefont {Pedersen}\ \emph {et~al.}(2012)\citenamefont
  {Pedersen}, \citenamefont {Gunst}, \citenamefont {Markussen},\ and\
  \citenamefont {Pedersen}}]{Pedersen2012}%
  \BibitemOpen
  \bibfield  {author} {\bibinfo {author} {\bibfnamefont {J.~G.}\ \bibnamefont
  {Pedersen}}, \bibinfo {author} {\bibfnamefont {T.}~\bibnamefont {Gunst}},
  \bibinfo {author} {\bibfnamefont {T.}~\bibnamefont {Markussen}}, \ and\
  \bibinfo {author} {\bibfnamefont {T.~G.}\ \bibnamefont {Pedersen}},\ }\href
  {\doibase 10.1103/PhysRevB.86.245410} {\bibfield  {journal} {\bibinfo
  {journal} {Phys. Rev. B}\ }\textbf {\bibinfo {volume} {86}},\ \bibinfo
  {pages} {245410} (\bibinfo {year} {2012})}\BibitemShut {NoStop}%
\bibitem [{\citenamefont {Gunst}\ \emph {et~al.}(2011)\citenamefont {Gunst},
  \citenamefont {Markussen}, \citenamefont {Jauho},\ and\ \citenamefont
  {Brandbyge}}]{Gunst2011}%
  \BibitemOpen
  \bibfield  {author} {\bibinfo {author} {\bibfnamefont {T.}~\bibnamefont
  {Gunst}}, \bibinfo {author} {\bibfnamefont {T.}~\bibnamefont {Markussen}},
  \bibinfo {author} {\bibfnamefont {A.-P.}\ \bibnamefont {Jauho}}, \ and\
  \bibinfo {author} {\bibfnamefont {M.}~\bibnamefont {Brandbyge}},\ }\href
  {\doibase 10.1103/PhysRevB.84.155449} {\bibfield  {journal} {\bibinfo
  {journal} {Phys. Rev. B}\ }\textbf {\bibinfo {volume} {84}},\ \bibinfo
  {pages} {155449} (\bibinfo {year} {2011})}\BibitemShut {NoStop}%
\end{thebibliography}
%

\end{document}